\documentclass[preprint,5p,times,twocolumn,authoryear]{elsarticle}
\usepackage{amsmath,amssymb,amsfonts} 
\usepackage{algorithmic} 
\usepackage{algorithm} 
\usepackage{graphicx} 
\usepackage{textcomp} 
\usepackage{xcolor} 
\usepackage{url} 
\usepackage{verbatim} 
\usepackage{multirow} 
\usepackage{booktabs} 
\usepackage{caption} 
\usepackage{subcaption} 
\usepackage{soulutf8} 
\usepackage{balance} 
\usepackage[hidelinks]{hyperref} 
\usepackage{diagbox} 
\usepackage{marvosym}
\usepackage{lineno}

\journal{Expert Systems with Applications}

\begin{document}

\begin{frontmatter}

\title{SIGNL: A Label-Efficient Audio Deepfake Detection System via Spectral-Temporal Graph Non-Contrastive Learning} 

\author[inst1,inst2]{Falih Gozi Febrinanto~\corref{cor1}} 
\ead{f.febrinanto@federation.edu.au}

\author[inst2]{Kristen Moore} 
\ead{kristen.moore@data61.csiro.au}

\author[inst2]{Chandra Thapa} 
\ead{chandra.thapa@data61.csiro.au}

\author[inst1]{Jiangang Ma} 
\ead{j.ma@federation.edu.au}

\author[inst1]{Vidya Saikrishna} 
\ead{v.saikrishna@federation.edu.au}

\affiliation[inst1]{organization={Institute of Innovation, Science and Sustainability, Federation University Australia},
            city={Ballarat},
            state={Victoria},
            country={Australia}}
            
\affiliation[inst2]{organization={CSIRO's Data61},
            country={Australia}}

\cortext[cor1]{Corresponding author}

\begin{abstract}
\textcolor{black}{Audio deepfake detection is increasingly important as synthetic speech becomes more realistic and accessible. Recent methods, including those using graph neural networks (GNNs) to model frequency and temporal dependencies, show strong potential but need large amounts of labeled data, which limits their practical use. Label-efficient alternatives like graph-based non-contrastive learning offer a potential solution, as they can learn useful representations from unlabeled data without using negative samples. However, current graph non-contrastive approaches are built for single-view graph representations and cannot be directly used for audio, which has unique spectral and temporal structures. Bridging this gap requires dual-view graph modeling suited to audio signals.
In this work, we introduce SIGNL (Spectral-temporal vIsion Graph Non-contrastive Learning), a label-efficient expert system for detecting audio deepfakes. SIGNL operates on the visual representation of audio—such as spectrograms or other time-frequency encodings—transforming them into spectral and temporal graphs for structured feature extraction. It then employs graph convolutional encoders to learn complementary frequency-time features, effectively capturing the unique characteristics of audio. These encoders are pre-trained using a non-contrastive self-supervised learning strategy on augmented graph pairs, enabling effective representation learning without labeled data. The resulting encoders are then fine-tuned on minimal labelled data for downstream deepfake detection.
SIGNL achieves strong performance on multiple audio deepfake detection benchmarks, including 7.88\% EER on ASVspoof 2021 DF and 3.95\% EER on ASVspoof 5 using only 5\% labeled data. It also generalizes well to unseen conditions, reaching 10.16\% EER on the In-The-Wild dataset when trained on CFAD. The code is available at \url{https://github.com/falihgoz/SIGNL}.
}
\end{abstract}


\end{frontmatter}

\section{Introduction} 
Audio deepfakes---synthetic speech generated using text-to-speech (TTS)~\citep{tan2024naturalspeech_TTS,wang2024spikevoice_TTS} or voice conversion technologies~\citep{choi2024dddm_vc,popov2021diffusion_vc}---are increasingly being used to impersonate individuals, commit fraud, and bypass automatic speaker verification (ASV) systems~\citep{yi2023audio_audioDeepfakeSurvey}. These threats pose serious challenges to voice-based authentication, customer service infrastructure, and media trust. As audio synthesis techniques rapidly improve, expert systems that can reliably detect audio deepfakes under practical constraints are urgently needed.

\textcolor{black}{Deep learning techniques such as convolutional neural networks (CNNs)~\citep{wu2020light_LCNN}, including advanced architectures like deep residual networks (ResNets)~\citep{wang2024multi_senet, tak2021end_cnnbased, chen2021pindrop_resnet}, as well as transformers~\citep{liu2023leveraging_transformer}, have been widely adopted for audio deepfake detection by adapting their architectures to process audio data or its visual representations, such as spectrograms. Graph Neural Networks (GNNs)~\citep{xia2025graph} have emerged as a promising alternative~\citep{chen2023graph_graph,jung2022aasist_graph}, offering the ability to model frequency sub-bands and temporal segments as graph structures. This enables GNNs to capture irregular and complex relationships, facilitating the detection of subtle deepfake artifacts that may span multiple frequency regions and time steps~\citep{han2022vision,chen2023graph_graph,jung2022aasist_graph}.}

\textcolor{black}{Despite their effectiveness, GNN-based methods for audio deepfake detection face major challenges because they depend heavily on labeled data~\citep{chen2023graph_graph, jung2022aasist_graph}. Large amounts of unlabeled audio, including call recordings, podcasts, and social media clips, are widely available but often remain unused due to the lack of ground truth labels. In situations with few labels, GNN-based methods tend to perform poorly and struggle to handle new attack types or unfamiliar domains~\citep{liu2022graph_graphssl}. Augmentations for raw audio, such as adding background noise or codec transformations~\citep{yamagishi2021asvspoof_21}, also tend to have limited effect on graph-level structures, making it difficult to enhance model robustness without labels~\citep{ding2022data_graphaug}. This presents a serious challenge in scenarios where large volumes of unlabeled audio—such as call recordings, podcasts, or social media clips—are available but underutilized due to the absence of ground truth.}

\textcolor{black}{
\textbf{Limitations of Existing Solutions.} Self-supervised learning (SSL), particularly non-contrastive learning, offers a promising way to build effective audio deepfake detectors without relying on large labeled datasets. Non-contrastive learning has shown success in computer vision~\citep{grill2020bootstrap_byol,chen2021exploring_siamese} and graph representation~\citep{guo2024architecture_graphclsurvey, ju2024towards_gcl}, where models learn from augmented positive pairs without using negative samples. This setting is useful in scenarios where false negatives can damage representation learning~\citep{liu2022discovering_falsenegativepushapart}. Non-contrastive SSL methods have also been applied directly to audio data. For example, BYOL-A~\citep{niizumi2022byol_byola} is proposed to model audio through its visual representation, such as spectrograms. However, applying non-contrastive learning to audio-derived graphs instead of raw audio or visual representations like spectrograms is still underexplored. This introduces challenges in how to structure, augment, and represent these graphs. Moreover, \textit{existing graph non-contrastive learning methods~\citep{guo2024architecture_graphclsurvey} typically work on a single graph view per instance and do not consider the distinct characteristics of audio data, such as the separation between spectral and temporal information}~\citep{ding2022data_graphaug}.
}

\textcolor{black}{\textbf{Our Work.} To address these limitations, we propose SIGNL (Spectral-temporal vIsion Graph Non-contrastive Learning), a practical expert system for label-efficient audio deepfake detection. Unlike existing graph non-contrastive learning methods that operate on single-view graphs and overlook the unique characteristics of audio, SIGNL constructs two complementary graphs—one modeling spectral dependencies,  the other capturing temporal relationships. \textit{These dual-view graphs more accurately reflect the distinct properties of audio signals}. SIGNL applies graph convolutional encoders trained using a non-contrastive self-supervised learning strategy, which maximizes the similarity between augmented spectral-temporal graph views. This enables the model to learn robust, generalizable representations, which can then be fine-tuned with minimal labelled data for downstream detection.}

We evaluate SIGNL on four benchmark datasets: ASVspoof 2021 DF~\citep{yamagishi2021asvspoof_21}, ASVspoof 5~\citep{yamagishi2021asvspoof_21}, CFAD~\citep{ma2024cfad}, and In-the-Wild~\citep{muller2022does_itw}. Results show that SIGNL consistently performs better than supervised models such as LCNN~\citep{wu2020light_LCNN} and AASIST~\citep{tak2022wav2vec_aasist}, and also outperforms self-supervised baselines including BYOL-A~\citep{niizumi2022byol_byola}, MoCo~\citep{he2020momentum_MoCo}, and a vanilla graph-based non-contrastive learning method used as a standard for graph non-contrastive learning~\citep{guo2024architecture_graphclsurvey}. Notably, SIGNL maintains strong performance even with only 5\% labeled data, demonstrating its suitability for deployment in real-world, data-constrained environments. \textbf{Our contributions} are summarized as follows:

\begin{itemize}
\item We propose SIGNL, a label-efficient expert system for audio deepfake detection, designed for practical use in settings with limited annotated data. 

\item We introduce a dual-graph construction strategy that captures both spectral and temporal structures from audio, enabling rich representation learning through label-free non-contrastive pre-training.

\item We demonstrate SIGNL's effectiveness across four benchmarks, showing superior performance to state-of-the-art methods in both low-label and cross-domain scenarios. 

\end{itemize}

The rest of this paper is organized as follows. Section~\ref{sec:related} gives an overview of related works, including current graph-based models for audio forensics, self-supervised learning for audio representation, and graph-based non-contrastive learning. Section~\ref{sec:preliminaries} introduces the notations and the problem we aim to solve. Section~\ref{sec:method} describes the details of our SIGNL for audio deepfake detection. We present the experimental results and discussion in Section~\ref{sec:experiment} and conclude the work in Section~\ref{sec:conclusion}.

\section{Related Work}
\label{sec:related}

\subsection{Graph-Based Models for Audio Forensics}
\textcolor{black}{GNNs, sometimes referred to as deep neural networks for graphs (DNNGs), are a subclass of deep learning designed to process graph-structured data. They provide a powerful framework for modeling complex relationships by representing data as nodes and edges and enabling the learning of structural patterns and relational dependencies~\citep{li2024guest}. Due to their flexibility and expressive capability, GNNs have been successfully applied across a wide range of application domains~\citep{xia2025graph}.}

Recently, GNNs have emerged as effective models for audio deepfake detection tasks, consistently outperforming traditional deep learning approaches~\citep{chen2023graph_graph,jung2022aasist_graph}. Techniques such as GAT-T~\citep{tak2021graph_GAT} and RawGAT-ST~\citep{tak2021end_rawGat} generate latent node embeddings from raw audio data using ResNet~\citep{he2016deep_restnetori} and GAT~\citep{velivckovic2018graph_GAT} to learn graph structures. The AASIST model~\citep{jung2022aasist_graph} further enhances this approach through heterogeneous stacking graph attention layers. Similarly,~\cite{chen2023graph_graph} create graph representations by segmenting spectrograms into patches, using linear filter banks (LFB) features to define nodes and edges, and processing the graphs with GCN~\citep{welling2016semi_gcn}. Recent hybrid methods integrate GNN classifiers with pre-trained self-supervised learning (SSL) models such as Wav2Vec2~\citep{baevski2020wav2vec} and Whisper~\citep{radford2023robust_whisper}, further improving detection accuracy~\citep{tak2022wav2vec_aasist,zhang2024audio_hybridlearnablevisual}.

Although these models achieve competitive performance in supervised settings, they are often not robust in label-scarce conditions. Their reliance on labels during graph construction and training limits their adaptability to diverse or evolving attack types. Furthermore, traditional audio augmentations such as noise addition or codec transformations~\citep{yamagishi2021asvspoof_21}do not meaningfully alter graph structure, offering limited benefits for improving GNN-based models in low-resource scenarios~\citep{ding2022data_graphaug}. This motivates the need for self-supervised strategies that can learn diverse, structure-aware representations without extensive supervision.

\begin{figure*}[!ht]
  \begin{center}
    \includegraphics[width=\textwidth]{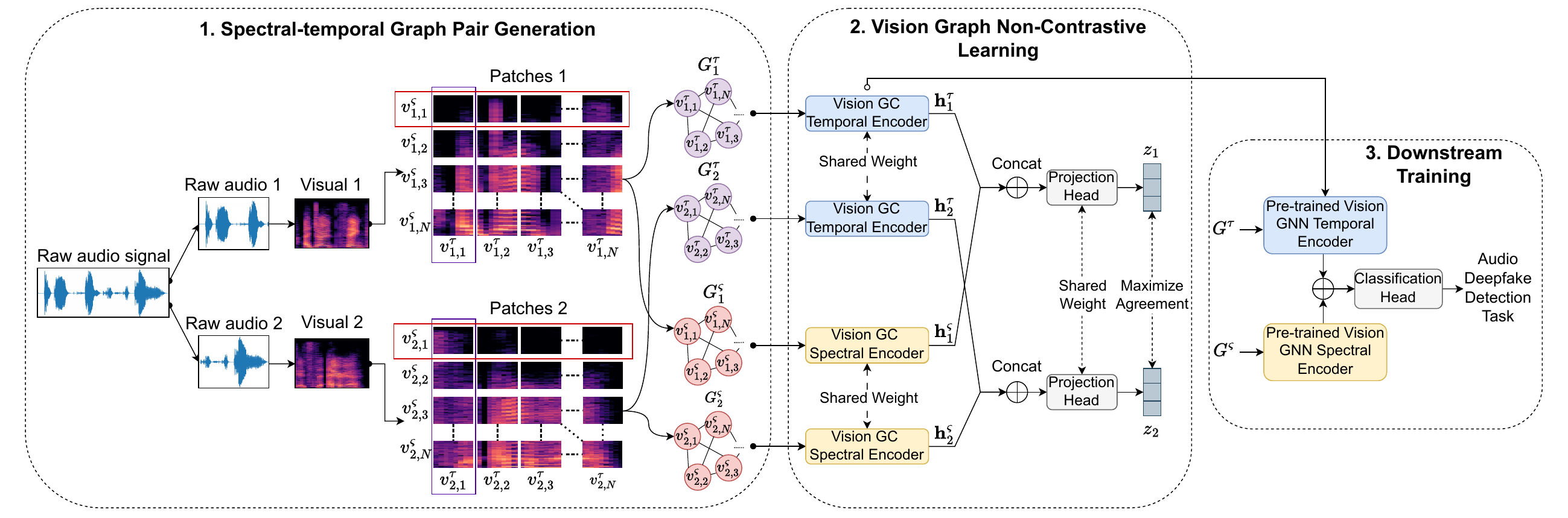}
  \end{center}
  \caption{\textcolor{black}{Overview of the SIGNL framework. \textbf{Stage 1 - Spectral-temporal graph generation:} Audio data is converted into spectral and temporal graphs to capture complementary frequency-time structure. \textbf{Stage 2 - Non-contrastive graph pre-training:} – Graph encoders are trained to maximize similarity between augmented graph pairs without labels. \textbf{Stage 3 - Downstream training:} – Pre-trained encoders are fine-tuned on a small labeled set for audio deepfake detection.}}
  \label{img:fram}
  \vspace{-2pt}
\end{figure*}

\subsection{Self-Supervised Learning for Audio Representation}
Self-supervised learning (SSL), particularly contrastive learning, has shown great promise in representation learning by encouraging similarity between augmented positive pairs and dissimilarity among negative samples~\citep{ericsson2022self_SSLsurvey}.
Prominent contrastive frameworks from computer vision such as SimCLR~\citep{chen2020simple_simclr}, MoCo~\citep{he2020momentum_MoCo}, BYOL~\citep{grill2020bootstrap_byol}, and SimSiam~\citep{chen2021exploring_siamese}, have been successfully adapted for audio applications. For example,
SimCLR~\citep{chen2020simple_simclr} was adapted to audio by~\cite{zhang2021contrastive_simclr}, MoCo~\citep{he2020momentum_MoCo} was utilized for speaker embedding~\citep{xia2021self_mocoEmbedding}, and COLA employs contrastive learning for universal audio understanding~\citep{saeed2021contrastive_COLA}.~\cite{wang2023self_acl} introduced Angular Contrastive Loss (ACL) to enhance the discriminative power of audio contrastive learning. Meanwhile, non-contrastive SSL methods, such as BYOL-A~\citep{niizumi2022byol_byola} learn from positive pairs alone, avoiding the sensitivity to false negatives that can hinder contrastive learning~\citep{ericsson2022self_SSLsurvey}.

While effective for raw audio or spectrogram inputs, these methods do not leverage the structured relational information present in graph-based representations derived from audio data.  As audio deepfake detection often depends on subtle and distributed artifacts, there is a growing need for domain-specific adaptations that enable SSL methods to harness graph-structured inputs that better reflect the relational nature of audio data.

\subsection{Graph-Based Non-Contrastive Learning}

\textcolor{black}{GNNs in real-world applications face several data-related challenges that can degrade their performance, including modeling heterophilous graph structures~\citep{zheng2022graph, li2024permutation}, noisy connectivity~\citep{zhou2025data}, and limited labeled data~\citep{liu2022graph_graphssl}. While structural issues affect graph design, label scarcity remains a major hurdle in practical applications where data annotation is costly. This issue is particularly relevant in audio deepfake detection, where large-scale unlabeled audio data are widely available but reliable labels are expensive to obtain, which limits the effectiveness of GNN-based models.
Graph self-supervised learning, including contrastive and non-contrastive learning, offers an effective solution, where graph non-contrastive learning enables representation learning from augmented positive graph pairs without relying on negative samples, thus reducing sensitivity to false negatives~\citep{liu2022discovering_falsenegativepushapart}.}

Inspired by BYOL~\citep{grill2020bootstrap_byol}, frameworks like BGRL~\citep{thakoorlarge_bgrl} introduced asymmetric architectures to avoid feature collapse.~\cite{zhang2021canonical_CCASSG} proposed canonical correlation analysis-based feature decorrelation, while Graph Barlow Twins (GBT)\citep{bielak2022graph_GBT} applied redundancy reduction principles to learn effective graph representations. However, most existing graph non-contrastive approaches target node classification tasks and operate on a single graph per instance.

In contrast, audio signals possess an inherently multi-faceted structure—spanning both spectral and temporal domains—that is not well captured by single-view graph models. Our framework addresses this gap by introducing a dual-view graph learning approach that explicitly models spectral and temporal graphs as separate but complementary inputs. This enables the system to integrate relational and temporal dynamics more effectively and supports robust audio deepfake detection under minimal supervision.

\section{Preliminaries}
\label{sec:preliminaries}

In this section, we first introduce basic concepts of managing graph structures derived from audio data and provide a high-level overview of our pre-trained vision GC encoders used in audio deepfake detection. Additionally, we summarize commonly used notations in Table~\ref{tab:notation}.

\begin{table}[!ht]
\small
\centering
\caption{Summary of Key Notations}
\label{tab:notation}
\resizebox{\columnwidth}{!}{%
\begin{tabular}{c|l}
\hline
\textbf{Notations} & \textbf{Description} \\
\hline
$G$ & Graph with nodes $V$, edges $E$, features $X$ \\
$V$ & Set of nodes in a graph \\
$E$ & Set of edges between nodes \\
$X$ & Node feature matrix \\
$e_{i,j}$ & Edge between nodes $v_i$ and $v_j$ \\
$x_i$ & Feature vector for node $v_i$ \\
$t$ & Audio clip duration in seconds \\
$F$ & Frequency dimension of audio’s visual representation \\
$T$ & Temporal dimension of audio's visual representation \\
$G^\tau$ & Temporal graph\\
$G^\varsigma$ & Spectral graph\\
$f^\tau$ & Temporal graph encoder \\
$f^\varsigma$ & Spectral graph encoder \\
$H^{(\ell)}$ & Set of node embeddings in layer $\ell$ \\
$h^{(\ell)}$ & A node embedding in layer $\ell$ \\
$\textbf{h}^\tau$ & Graph-level representation for temporal graph\\
$\textbf{h}^\varsigma$ & Graph-level representation for spectral graph\\
$z$ & Vector representation of an audio \\
$s(z_1, z_2)$ & Cosine similarity of representations $z_1, z_2$ \\
$g$ & Projection head for dimensional reduction \\
$c$ & Classification head for the downstream task \\
$y$ & Classification output\\
$\mathcal{L}_{align}$ & Alignment loss for positive pairs \\
$\mathcal{L}_{CE}$ & Cross-entropy loss for detection training \\
\hline
\end{tabular}
}
\end{table}

\noindent \textbf{Graph Data.} In this work, a graph is formally defined as $G = (V, E, X)$
where $V = \{v_1, v_2, \dotsc, v_N\}$ represents
a set of nodes in a graph with $N$ nodes. For $i\neq j$ and $i,j\in \{1,2,\dotsc,N\}$, each edge $e_{i,j} = (v_i, v_j) \in E$ indicates a relationship between two nodes. The feature vector $x_i \in X$ is associated with each node $v_i$, where $X \in \mathbb{R}^{N \times D}$, and $D$ denotes the dimension of the feature vector.

\noindent \textbf{Spectral-temporal Graphs of Audio.} In SIGNL, each iteration of non-contrastive learning generates two audio samples as positive pairs. Each sample is represented by two types of graphs: a temporal graph $G^\tau$ and a spectral graph $G^\varsigma$. For the first audio sample, this produces the temporal graph $G^\tau_1 = (V^\tau_1, E^\tau_1, X^\tau_1)$ and the spectral graph $G^\varsigma_1 = (V^\varsigma_1, E^\varsigma_1, X^\varsigma_1)$, and for the second audio sample, this produces the temporal graph $G^\tau_2 = (V^\tau_2, E^\tau_2, X^\tau_2)$ and the spectral graph $G^\varsigma_2 = (V^\varsigma_2, E^\varsigma_2, X^\varsigma_2)$, where the indices 1 or 2 denote the corresponding audio sample. Given the temporal graph $G^\tau_1 = (V^\tau_1, E^\tau_1, X^\tau_1)$, $V^\tau_1 = \{v^\tau_{1,1}, v^\tau_{1,2}, \dotsc, v^\tau_{1,N}\}$ represents the set of nodes from the temporal graph of audio sample 1, where $N$ denotes the total number of nodes.

\noindent \textbf{Pre-trained Vision GC Encoders.} SIGNL produces pre-trained encoders through a graph non-contrastive learning mechanism. Two encoders are pre-trained: a vision GC temporal encoder $f^\tau$ that is trained on the input temporal graph pair $G^\tau_1$ and $G^\tau_2$, and a vision GC spectral encoder $f^\varsigma$ that is trained on the input spectral graph pair $G^\varsigma_1$ and $G^\varsigma_2$. The pre-trained encoders $f^\tau$ and $f^\varsigma$ are then fine-tuned for the downstream task of audio deepfake detection. This combination of encoders generates a generic representation $z$, which is passed through a classification head to produce a prediction: $y=1$ for genuine (bona fide) audio or $y=0$ for fake audio.

\section{Methodology}
\label{sec:method}
The SIGNL framework for audio deepfake detection (illustrated in \autoref{img:fram}) is composed of three main components: \textit{spectral-temporal graph pair generation}, \textit{vision graph non-contrastive learning}, and \textit{downstream training}. Each stage is designed to address challenges specific to low-label learning in graph-based deepfake detection. In this section, we describe each component in detail.

\subsection{Spectral-temporal Graph Pair Generation}
This section explains the components of managing positive temporal and spectral graph pairs from audio data. The steps include generating positive pairs, converting audio data into a graph structure, and performing graph data augmentation.

\noindent \textbf{Generating Positive Pairs of Audio.} Our work adopts a non-contrastive learning approach, which excludes negative samples, distinguishing it from traditional contrastive learning methods. In contrastive frameworks such as COLA~\citep{saeed2021contrastive_COLA}, treating other audio samples in the same batch as negative pairs can lead to unintended outcomes. Specifically, samples from the same class--such as \textit{genuine} or \textit{fake}--may be incorrectly pushed apart in the feature space due to the presence of false negatives~\citep{liu2022discovering_falsenegativepushapart}. Moreover, recent work by~\cite{guo2024architecture_graphclsurvey} suggests that negative pairs are unnecessary for graph contrastive learning for graph classification tasks and that comparable performance can be achieved without them.

To generate positive pairs of raw audio, we adopt a lightweight audio segmentation strategy inspired by COLA~\citep{saeed2021contrastive_COLA}. This involves splitting the original audio into two segments of similar size, ensuring consistent positive pairs while enabling efficient processing. Each audio clip is standardized to a fixed duration of $t$ seconds. For clips shorter than $t$, the audio is repeated to meet the required length, whereas longer clips are truncated. Positive pairs are then sampled from the same $t$-second audio, producing two segments: audio sample 1 and audio sample 2, each with a duration of $\left\lfloor \frac{t}{2} \right\rfloor$ seconds. Each segment is then transformed into a visual representation---such as a spectrogram---that forms the basis for graph construction.

\noindent \textbf{Converting Audio Pairs into Spectral-temporal Graphs}. To convert audio data into a graph structure, we divide its visual representation, such as a spectrogram, into patches. A 2D visual representation of size $F \times T$---where $F$ denotes the frequency axis and $T$ the temporal axis---is divided into $N$ patches along both the horizontal (temporal segmentation) and vertical (spectral segmentation) dimensions. Each temporal patch is transformed into a feature vector $x^\tau_i \in \mathbb{R}^D$, forming the overall feature matrix $X^\tau = [x_1^\tau, x_2^\tau, \dots, x_N^\tau]$. Similarly, each spectral patch is transformed into a feature vector $x^\varsigma_i \in \mathbb{R}^D$, with the overall feature matrix represented as $X^\varsigma = [x_1^\varsigma, x_2^\varsigma, \dots, x_N^\varsigma]$. 

These feature vectors correspond to nodes, denoted as $V^\tau$ for temporal patches and $V^\varsigma$ for spectral patches. For each node $v_i$ in both views, we identify its $K$ nearest neighbors $\mathcal{N}(v_i)$ based on the Euclidean distance and establish edges $e_{ji}$ from neighboring nodes $v_j$ to $v_i$ for all $v_j \in \mathcal{N}(v_i)$. This process results in  fully constructed spectral-temporal graphs $G^\tau = (V^\tau, E^\tau, X^\tau)$ for the temporal view and $G^\varsigma = (V^\varsigma, E^\varsigma, X^\varsigma)$ for the spectral view. Each step processes two audio samples as a positive pair, yielding four graphs in total: $G^\tau_1$ and $G^\varsigma_1$ for audio sample 1, and $G^\tau_2$ and $G^\varsigma_2$ for audio sample 2.

\noindent \textbf{Graph-based Augmentations}. This process modifies the spectral-temporal graphs in the positive pairs by dynamically introducing varied graph characteristics during the pre-training phase. Diversification of graph properties enables the model to better capture underlying structures, enhances robustness to noise, and improves generalization to unseen graph structures, thereby enhancing performance in downstream tasks~\citep{ju2024towards_gcl}. Building on the work of~\cite{guo2024architecture_graphclsurvey}, we employ three types of graph-based augmentations:

\begin{itemize}
\item \textbf{Edge Dropping (ED)}: Randomly removes edges from the graph, increasing sparsity.
\item \textbf{Gaussian Noise (GN)}: Adds random noise sampled from a Gaussian distribution to the node features.
\item \textbf{Feature Masking (FM)}: Randomly sets a proportion of node features to zero.
\end{itemize}

The optimal combination of these graph augmentations is analyzed in \autoref{sec:graph_aug}. 

\begin{figure}[!ht]
  \centering
  \includegraphics[width=0.9\columnwidth]{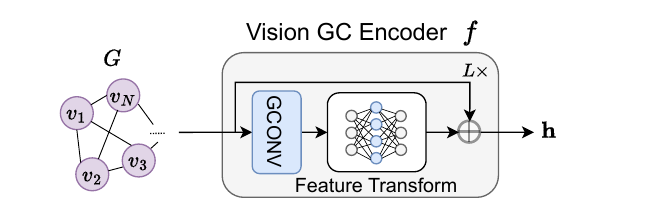}
  \caption{Vision GC Encoder.}
  \label{img:vision}
  \vspace{-2pt}
\end{figure}

\subsection{Vision Graph Non-Contrastive Learning}
This section introduces our vision graph non-contrastive learning module, which processes positive pairs composed of two graphs, a temporal graph and a spectral graph, extracted from each sample. Unlike existing graph non-contrastive learning approaches~\citep{guo2024architecture_graphclsurvey}, which operate on a single graph per instance and overlook the specific characteristics of each view, SIGNL explicitly models and fuses both spectral and temporal graph views to preserve the unique structure of audio data. Fusing spectral and temporal graphs captures correlations across frequencies in the spectral graph and dependencies that evolve over time in the temporal graph at the same time. The vision graph non-contrastive learning module enables the pre-training process to learn graph representations without labels by maximizing the similarity between positive graph pairs.

\noindent \textbf{Vision Graph Convolution Encoder}. The Vision GC encoder, inspired by Vision GNN~\citep{han2022vision}, was originally developed for image classification and object detection tasks. \textit{We use the term ``vision" in this work because we model the visual representation of audio, such as spectrograms or other 2D matrices that can be visualized as heatmaps}. We further extend this technique to make it suitable for processing audio data. After deriving graph structures from the audio, each graph $G$ is processed through the vision GC encoder. The GNNs model graph representation by enabling nodes to exchange information through feature aggregation from their neighbors, where nodes represent patches in the audio's visual representation. 

The proposed vision GC encoder is depicted in~\autoref{img:vision}. This encoder models a graph $G$ through a function defined as vision GC $f$, producing an output $\textbf{h}$, expressed as $f(G) \rightarrow \textbf{h}$. For the encoder's first component, we employ a GCN~\citep{welling2016semi_gcn} due to its simplicity and efficiency. The GCN layer is defined as follows:

\begin{equation}
\label{eq:gcn}
H^{(\ell+1)} = \sigma (\hat{D}^{-\frac{1}{2}}\hat{A}\hat{D}^{-\frac{1}{2}}H^{(\ell)}\theta^{(\ell)}),
\end{equation}
where $H^{(\ell+1)}$ represents the updated node feature matrix, and $H^{(\ell)}$ represents the node embeddings from the previous layer. The initial node embeddings $H^0$ correspond to the original node feature matrix $X$ of graph $G$. Here $\hat{A} = A + I$ denotes the adjacency matrix with self-loops, $\hat{D}$ is the diagonal degree matrix of $\hat{A}$, and $\theta^{(\ell)}$ are the learnable parameters of layer $\ell$. The function $\sigma(\cdot)$ is a nonlinear activation function (\emph{e.g.}, ReLU).

We implement the multi-head operation as a \textit{feature transformation module} to enhance the expressiveness of node embeddings after graph convolution by learning different feature components in each node embedding. The transformation involves splitting the feature embedding of each node $h^{(\ell+1)}_i$ into $K$ heads:  $head_1, head_2, \ldots, head_K$. Each head projects the node representation using different weights. The feature embeddings from all heads are updated in parallel, and their results are concatenated to form the updated feature vector:

\begin{equation}
\begin{aligned}
\label{eq:multiHead}
h^{(\ell+1)}_i &= \mathbin\Vert_{j=1}^K head_j \theta_j \\
&= [head_1 \theta_1 \Vert head_2 \theta_2 \Vert \ldots \Vert head_K \theta_K].
\end{aligned}
\end{equation}
This multi-head operation improves the model's feature diversity. To address the vanishing gradient problem, we incorporate a skip connection in each layer of the GNN encoder. The updated feature embedding of each node, which combines the skip connection with the multi-head operation, is given by $h^{(\ell+1)}_i = \sigma(h_i^{(\ell+1)}) + h_i^{(\ell)}$, where $\sigma(\cdot)$ denotes an activation function (\emph{e.g.}, ReLU), and $h^{(\ell)}$ represents the node's previous representation before the GCN operation.

In this study, the node embeddings $H$ after $\ell$ layers of convolution are concatenated to form the graph-level representation by aggregating nodes. For each graph, CONCAT pooling is performed as follows:

\begin{equation}
\label{eq:concat}
\textbf{h} = \mathbin\Vert_{j=1}^N h_j = [h_1 \Vert h_2 \Vert \ldots \Vert h_N].
\end{equation}

During the vision graph non-contrastive learning process, we pre-train two encoders through an unsupervised, label-free mechanism. The temporal encoder $f^\tau$ is trained on the temporal graph pair $G^\tau_1$ and $G^\tau_2$ using shared weights, meaning the same encoder is applied to both graphs. Similarly, the spectral encoder $f^\varsigma$ is trained separately on the spectral graph pair $G^\varsigma_1$ and $G^\varsigma_2$, also with shared weights.

\noindent \textbf{Projection Head}. In typical graph non-contrastive learning, the projection head processes a single graph representation obtained after graph embedding~\citep{guo2024architecture_graphclsurvey}. However, since both spectral and temporal aspects play a crucial role in audio data~\citep{tak2022wav2vec_aasist}, our approach introduces a mechanism to fuse information from both spectral and temporal graphs, derived from the visual representation of audio data, before passing them into the projection head of the graph non-contrastive learning framework.

Each audio sample in a positive pair generates spectral-temporal graph representations: $ \textbf{h}^\tau$ for the temporal graph and $\textbf{h}^\varsigma$ for the spectral graph. These representations are jointly fused via concatenation and passed through a shared projection head $ g $, which reduces dimensionality to produce a unified vector representation $ z = g(\textbf{h}^\tau \Vert \textbf{h}^\varsigma) $. This fusion captures both spectral and temporal information, ensuring that both spectral and temporal cues are integrated within the learned embeddings. As a result, two vectors are obtained for the positive pair: $ z_1 $ for audio sample 1 and $ z_2 $ for audio sample 2. The projection head, implemented as a simple multi-layer perceptron with shared weights, is applied identically to both positive samples.

\noindent \textbf{Alignment Loss}.
In contrastive learning, the InfoNCE loss~\citep{oord2018representation_infoNCE} is typically used to bring positive samples closer together and separate negative samples. However, since our framework consists only of positive pairs, we adopt the alignment component of the InfoNCE loss, denoted as $\mathcal{L}_{align}(z_1,z_2)$. This approach prioritizes maximizing the alignment between positive graph pairs, without requiring contrastive separation from negative samples. Given a positive pair of vector representations $z_1$ and $z_2$, the alignment loss is defined as:

\begin{equation}
    \label{eq:align}
\mathcal{L}_{align}(z_1,z_2) = -\sum \frac{s(z_1, z_2)}{t},
\end{equation}
where $s(z_1, z_2) = \frac{z_1^\top z_2}{\|z_1\| \|z_2\|}$ is the cosine similarity between the positive samples, and $t$ is a temperature parameter that scales the similarity scores. Optimizing this loss increases the similarity for positive pairs. The complete algorithm for spectral-temporal vision graph non-contrastive learning is described in Algorithm~\autoref{al:const}. 

It is important to note that optimizing for positive similarity can result in the projection head $g$ learning constant features for both $z_1$ and $z_2$, a phenomenon known as feature collapse~\citep{guo2024architecture_graphclsurvey}. However, while the projection head $g$ may exhibit collapse, the vision GC encoders remain unaffected. The projection head $g$ introduces a non-linear transformation, ensuring that the encoders learn informative and distinct features~\citep{guo2024architecture_graphclsurvey}. Consequently, the pre-trained Vision GC encoders are effectively transferable for downstream tasks. Feature collapse is further discussed in~\autoref{sec:graph_aug}.

\begin{algorithm}[!ht]
  \caption{Vision Non-Graph Contrastive Learning.}
  \small
  \label{al:const}
  \begin{algorithmic}[1]
  \REQUIRE Training Data $\mathcal{D} = \{(G^\tau_{1}, G^\varsigma_{1}, G^\tau_{2}, G^\varsigma_{2})\}_{i=1}^M$, number of epochs $\text{Ep}$, batch size $\text{Bt}$.
  \ENSURE Pre-trained GNN encoders $f^\tau$ and $f^\varsigma$.
  \STATE \textbf{Parameters:} Temporal GNN encoder $f_\theta^\tau$, Spectral GNN encoder $f_\theta^\varsigma$, projection head $g_\phi$
  \FOR{epoch $= 1$ to $\text{Ep}$}
      \FOR{each batch $\{(G^\tau_{1}, G^\varsigma_{1}, G^\tau_{2}, G^\varsigma_{2})\}$ of size $\text{Bt}$ from $\mathcal{D}$}
          \STATE $G^\tau_1 \leftarrow \text{Augment}(G^\tau_{1})$ \COMMENT{Augment each $G^\tau_{1}$ in the batch}
          \STATE $G^\varsigma_1 \leftarrow \text{Augment}(G^\varsigma_{1})$ \COMMENT{Augment each $G^\varsigma_{1}$ in the batch}
          \STATE $G^\tau_2 \leftarrow \text{Augment}(G^\tau_{2})$ \COMMENT{Augment each $G^\tau_{2}$ in the batch}
          \STATE $G^\varsigma_2 \leftarrow \text{Augment}(G^\varsigma_{2})$ \COMMENT{Augment each $G^\varsigma_{2}$ in the batch}
          \STATE $\textbf{h}_1^\tau \leftarrow f_\theta^\tau(G^\tau_1)$ \COMMENT{Apply encoder to the graph $G^\tau_1$}
          \STATE $\textbf{h}_1^\varsigma \leftarrow f_\theta^\varsigma(G^\varsigma_1)$ \COMMENT{Apply encoder to the graph $G^\varsigma_1$}
          \STATE $\textbf{h}_2^\tau \leftarrow f_\theta^\tau(G^\tau_2)$ \COMMENT{Apply encoder to the graph $G^\tau_2$}
          \STATE $\textbf{h}_2^\varsigma \leftarrow f_\theta^\varsigma(G^\varsigma_2)$ \COMMENT{Apply encoder to the graph $G^\varsigma_2$}
          \STATE $z_1 \leftarrow g_\phi(\textbf{h}^\tau_1 \Vert \textbf{h}^\varsigma_1)$ \COMMENT{Project embeddings of the anchor}
          \STATE $z_2 \leftarrow g_\phi(\textbf{h}^\tau_2 \Vert \textbf{h}^\varsigma_2)$ \COMMENT{Project embeddings of the positive}
          \STATE $\mathcal{L}_{align} \leftarrow \mathcal{L}_{align}(z_1, z_2)$ \COMMENT{Calculate the loss}
          \STATE Update $\theta^\tau, \theta^\varsigma, \phi$ to minimize $\mathcal{L}_{align}$ \COMMENT{Backpropagation}
      \ENDFOR
  \ENDFOR
  \RETURN $f_\theta^\tau$, $f_\theta^\varsigma$ \COMMENT{Return the pre-trained encoders}
  \end{algorithmic}
\end{algorithm}

\subsection{Downstream Training}

Downstream training fine-tunes the pre-trained encoders for the audio deepfake detection task, addressing scenarios with \textit{limited labeled data}. Unlike a fixed approach that keeps parameters unchanged, we use a fine-tuning strategy to slightly adjust these encoder parameters for the new task. A comparison of the fixed and fine-tuning approaches is provided in~\autoref{sec:ablation}.

Similar to pre-training, downstream training uses temporal and spectral graphs, denoted as $G^\tau$ and $G^\varsigma$, derived from the audio data. These graphs are input to the pre-trained temporal GNN encoder $f^\tau$ and spectral GNN encoder $f^\varsigma$. However, in the downstream training,  the projection head $g$ is replaced with a classification head $c$, which avoids the potential issue of collapsed features generated by $g$. The classification head $c$, implemented as a multi-layer perceptron, is designed to adapt to the audio detection task. 

The graph-level representation $\textbf{h}^\tau$ and $\textbf{h}^\varsigma$ produced from the pre-trained temporal and spectral GNN encoders are concatenated and passed to the classification head $c$ to obtain the final prediction:
\begin{equation}
    y = c(\textbf{h}^\tau \Vert \textbf{h}^\varsigma),
\end{equation}
where $y = 1$ indicates genuine (bona fide) audio, and $y = 0$ indicates fake audio. Downstream training minimizes the cross-entropy loss $\mathcal{L}_{CE}(y, \hat{y})$, where $y$ is the predicted label,  and $\hat{y}$ is the true label. The loss function is defined as:

\begin{equation}
\label{eq:bce}
\mathcal{L}_{CE}(y, \hat{y}) = -\sum_{i=1}^{C} \hat{y}_i \log(y_i),
\end{equation}
where $C=2$ represents the number of classes: bona fide and fake. The complete algorithm for downstream training is described in Algorithm~\autoref{al:down}.

\begin{algorithm}[!ht]
    \caption{Downstream Training.}
    \small
    \label{al:down}
    \begin{algorithmic}[1]
    \REQUIRE Training Data $\mathcal{D} = \{(G^\tau, G^\varsigma, \hat{y})\}_{i=1}^M$ with labels, number of epochs $\text{Ep}$, batch size $\text{Bt}$.
    \ENSURE Trained model for audio deepfake detection tasks.
    \STATE \textbf{Parameter:} Classification head $c_\phi$.
    \STATE Load parameters for $f_\theta^\tau$ and $f_\theta^\varsigma$ from pre-trained models
    \FOR{epoch $= 1$ to $\text{Ep}$}
        \FOR{each batch $\{(G^\tau, G^\varsigma, \hat{y})\}$ of size $\text{Bt}$ from $\mathcal{D}$}
          \STATE $\textbf{h}^\tau \leftarrow f_\theta^\tau(G^\tau)$ \COMMENT{Apply encoder to each $G^\tau$ in the batch}
          \STATE $\textbf{h}^\varsigma \leftarrow f_\theta^\varsigma(G^\varsigma)$ \COMMENT{Apply encoder to each $G^\varsigma$ in the batch}
          \STATE $y \leftarrow c_\phi(\textbf{h}^\tau \Vert \textbf{h}^\varsigma)$ \COMMENT{Concatenate and classify embeddings}
          \STATE $\mathcal{L}_{CE} \leftarrow \mathcal{L}_{CE}(y, \hat{y})$ \COMMENT{Calculate the batch loss}
          \STATE Update $\theta^\tau$, $\theta^\varsigma$, $\phi$ to minimize $\mathcal{L}_{CE}$ \COMMENT{Backpropagation}
        \ENDFOR
    \ENDFOR
    \RETURN $f_\theta^\tau$, $f_\theta^\varsigma$, $c_\phi$ \COMMENT{Return the trained model}
    \end{algorithmic}
  \end{algorithm}
  
\subsection{Complexity Analysis}
The end-to-end audio deepfake detection pipeline comprises three main components: \emph{Spectral-temporal Graph Pair Generation}, \emph{Vision Graph Non-Contrastive Learning}, and \emph{Downstream Training}. Initially, each audio sample is converted into two graphs (temporal and spectral) by dividing its spectrogram into $N$ patches (nodes), each represented by a $D$-dimensional feature vector, resulting in a naive graph construction complexity of $O(N^2 \cdot D)$. Next, these graphs are processed through a vision GC encoder consisting of $L$ GCN layers, where within each layer, each node aggregates information from its $K$ nearest neighbors, incurring a complexity of $O(L \cdot N \cdot K \cdot D)$. For simplicity in this complexity analysis, we assume that the embedding dimension remains the same as the initial feature dimension $D$. Finally, a projection head or a classification head fuses the graph-level representations, adding minor computational overhead. Although pre-training involves twice as many graphs per sample (four graphs per positive pair) compared to downstream training (processing one temporal and one spectral graph per audio sample), both share the same asymptotic complexity of $O(N^2 \cdot D) + O(L \cdot N \cdot K \cdot D)$. Overall, SIGNL maintains manageable computational complexity while enabling expressive spectral-temporal representation learning with minimal supervision.

\section{Experiments}
\label{sec:experiment}
In this section, we comprehensively evaluate SIGNL to answer four key research questions related to detection performance, graph augmentation effectiveness, model design, and hyperparameter robustness.
\begin{itemize}
  \item \textbf{RQ1}: Can SIGNL outperform existing baselines in both \textit{in-domain} evaluation, where the training and evaluation sets originate from similar audio deepfake dataset sources, and \textit{cross-domain} evaluation, where the training and evaluation sets come from different dataset sources, even \textit{with limited labeled data?}
  \item \textbf{RQ2}: Do the three combinations of graph augmentations contribute significantly to SIGNL's performance?
  \item \textbf{RQ3}: What is the impact of each module in SIGNL on improving audio deepfake detection?
  \item \textbf{RQ4}: How does SIGNL's performance vary with different settings of key parameters?
\end{itemize}

\begin{table*}[!ht]
\centering
\caption{Statistics for training (train), development (dev), and evaluation (eval) sets across datasets.}
\label{tb:datasets}
\resizebox{\textwidth}{!}{%
\begin{tabular}{r|ccc|ccc|ccc|ccc}
\hline
\multicolumn{1}{c|}{\multirow{2}{*}{Set}} & \multicolumn{3}{c|}{ASVspoof 2021 DF (English)} & \multicolumn{3}{c|}{ASVspoof5 (English)} & \multicolumn{3}{c|}{CFAD (Chinese)} &  \multicolumn{3}{c}{In-The-Wild (English)} \\ \cline{2-13} 
\multicolumn{1}{c|}{} & Bona fide & Fake & Total & Bona fide & Fake & Total & Bona fide & Fake & Total & Bona fide & Fake & Total \\ \hline
Train & 2,580 & 22,800 & 25,380 & 18,797 & 163,560 & 182,357 & 12,800 & 25,600 & 38,400  & - & - & - \\
Dev & 2,548 & 22,296 & 24,844 & 15,667 & 41,106 & 56,773 & 4,800 & 9,600 & 14,400 & - & - & -  \\
Eval & 22,617 & 589,211 & 611,828 & 15,667 & 68,510 & 84,177 & 21,000 & 42,000 & 63,000  & 19,963 & 11,816 & 31,779 \\ \hline
\end{tabular}
}
\end{table*}

\subsection{Datasets and Evaluation Metrics}
\label{sec:datasets}
We evaluate SIGNL and baseline methods using publicly available and commonly used datasets: ASVspoof 2021 DF~\citep{yamagishi2021asvspoof_21}, the newly released ASVspoof 5~\citep{wang2024asvspoof_ASVspoof5}, and the Chinese Fake Audio Detection (CFAD)\citep{ma2024cfad} dataset under different limited data scenarios. We additionally use the In-The-Wild\citep{muller2022does_itw} dataset to assess performance on unseen data distributions.

\textbf{ASVspoof 2021 DF:} We use the evaluation set from the deepfake (DF) track, which focuses on audio deepfake detection under realistic conditions. The dataset contains bona fide and fake speech processed through various audio codecs. Its training and development sets are sourced from the logical access (LA) track of ASVspoof 2019~\citep{todisco2019asvspoof_asvspoof2019}, featuring fake audio generated through speech synthesis and voice conversion techniques.

\textbf{ASVspoof 5:} The latest ASVspoof challenge dataset includes crowdsourced data with diverse acoustic conditions. We use its training and development sets, which include both bona fide and attack samples. The training set contains attacks A01–A08, while the development set contains attacks A09–A16. Since the official test set labels have not yet been released, we simulate a test set by splitting the development set into two parts: one for validation (containing A09–A14) and another for evaluation (A09–A14 with additional unseen attacks, A15 and A16).

\textbf{CFAD:} The Chinese Fake Audio Detection (CFAD) dataset features Chinese speakers and includes four subsets: training, development, seen test, and unseen test. The unseen test contains new attack types and scenarios to evaluate model generalization. For our experiments, we combine the seen and unseen test subsets into a unified test set.

\textbf{In-The-Wild:} To evaluate robustness in unseen scenarios, we perform cross-domain evaluations using the In-the-Wild dataset~\citep{muller2022does_itw}, which includes audio deepfakes of politicians and public figures collected from platforms like video streaming and social media. Dataset statistics are summarized in \ref{tb:datasets}.

Importantly, we do not use extensive raw audio augmentations (e.g., codec conversion, reverberation, background noise), which are commonly used in prior work. Instead, our experiments focus on \textit{limited label scenarios} using the original unprocessed data without any pre-processing to ensure a fair comparison across all models in the same unaltered data conditions.

\textbf{EER Metric:} Performance is measured using the \textbf{equal error rate (EER)}, a standard metric in biometric systems. A lower EER indicates better performance in distinguishing bona fide from fake audio samples. We focus on EER as it is a standard and balanced metric in biometric security tasks, reflecting the trade-off between false acceptance and false rejection.

\subsection{Experimental Settings}
\label{sec:settings}
Each audio clip was processed to a uniform length of $t=9$ seconds. Shorter audio clips were repeated to meet this duration, while longer clips were trimmed. During pre-training, each positive pair consisted of two 4.5-second segments randomly extracted from the 9-second audio. In downstream training, only 4.5-second audio segments were used as inputs.

To transform raw audio data into a 2D matrix, we used Wav2Vec2~\citep{baevski2020wav2vec}, a pre-trained SSL model for automatic speech recognition (ASR). Specifically, the wav2vec2-xls-r-300m variant was employed to generate feature matrices serving as visual representations of the audio, fine-tuned for audio deepfake detection tasks. Each resulting 2D matrix had dimensions of (1024, 224). These matrices were divided into 32 temporal and 32 spectral patches, resulting in two graph views (temporal and spectral) with 32 nodes each across all datasets. During model training, we utilized a \textit{k-nearest neighbors} graph with three outgoing edges per node for ASVspoof 2021 DF and ASVspoof 5, and four for CFAD. Details on parameter sensitivity are provided in~\autoref{sec:sens}.

Experiments were conducted on an Intel Xeon Platinum 8452Y 3.20 GHz processor and an NVIDIA H100 GPU. The visual representations were segmented into patches using a \textit{stem convolutional layer}~\citep{han2022vision} before being passed into the vision GC encoders. Each node had a 256-dimensional feature vector. The vision GC encoders consisted of a 5-layer pyramid architecture for both temporal and spectral graph processing, halving the embedding size at each layer. A 3-layer projection head with 256, 128, and 80 hidden dimensions was used during pre-training, which was subsequently replaced by a classification head during the downstream training stage. 

Training was performed with a batch size of 96 using the Adam optimizer. The learning rates were set to 0.00001 for ASVspoof 2021 DF and ASVspoof 5, and 0.0001 for CFAD. Both pre-training and downstream training ran for 100 epochs with early stopping criteria applied.

\begin{table*}[htbp]
\centering
\caption{EER (\%) comparison across different training label percentages for in-domain evaluation. \textcolor{black}{All values of the methods\protect\footnotemark[2]} represent mean $\pm$ standard deviation. Lower values: better performance ($\downarrow$). \textbf{Bold}: best result.}
\label{tab:results}
\resizebox{\textwidth}{!}{%
\begin{tabular}{cllcccccc}
\hline
\multicolumn{1}{c|}{\multirow{2}{*}{\textbf{Categories}}} & \multicolumn{1}{c|}{\multirow{2}{*}{\textbf{Methods}}} & \multicolumn{1}{c|}{\multirow{2}{*}{\textbf{Front-end}}} & \multicolumn{6}{c}{\textbf{ASVspoof 2021 DF}} \\ \cline{4-9} 
\multicolumn{1}{c|}{} & \multicolumn{1}{c|}{} & \multicolumn{1}{c|}{} & \textbf{5\%} & \textbf{10\%} & \textbf{30\%} & \textbf{50\%} & \textbf{80\%} & \textbf{Full} \\ \hline
\multicolumn{1}{c|}{\multirow{6}{*}{Supervised}} & \multicolumn{1}{l|}{LCNN} & \multicolumn{1}{l|}{LFCC} & 30.26 $\pm$ 2.03 & 28.91 $\pm$ 2.57 & 25.63 $\pm$ 0.46 & 25.99 $\pm$ 0.72 & 25.21 $\pm$ 0.45 & 25.44 $\pm$ 0.82 \\
\multicolumn{1}{c|}{} & \multicolumn{1}{l|}{LCNN} & \multicolumn{1}{l|}{Whisper} & 24.61 $\pm$ 4.35 & 18.09 $\pm$ 1.85 & 23.49 $\pm$ 8.63 & 19.38 $\pm$ 2.43 & 20.32 $\pm$ 4.05 & 20.64 $\pm$ 5.90 \\
\multicolumn{1}{c|}{} & \multicolumn{1}{l|}{LCNN} & \multicolumn{1}{l|}{Wav2Vec2} & 38.42 $\pm$ 5.24 & 28.52 $\pm$ 7.56 & 9.36 $\pm$ 2.79 & 11.90 $\pm$ 3.90 & 8.63 $\pm$ 0.93 & 12.16 $\pm$ 3.06 \\
\multicolumn{1}{c|}{} & \multicolumn{1}{l|}{AASIST} & \multicolumn{1}{l|}{Raw Waveform} & 25.45 $\pm$ 3.88 & 23.72 $\pm$ 1.15 & 22.79 $\pm$ 1.76 & 22.78 $\pm$ 2.41 & 22.48 $\pm$ 1.33 & 22.85 $\pm$ 1.52 \\
\multicolumn{1}{c|}{} & \multicolumn{1}{l|}{AASIST} & \multicolumn{1}{l|}{Whisper} & 14.87 $\pm$ 1.19 & 14.55 $\pm$ 0.99 & 13.31 $\pm$ 0.47 & 14.06 $\pm$ 0.64 & 14.40 $\pm$ 1.05 & 16.48 $\pm$ 1.66 \\
\multicolumn{1}{c|}{} & \multicolumn{1}{l|}{AASIST} & \multicolumn{1}{l|}{Wav2Vec2} & 28.81 $\pm$ 8.72 & 8.47 $\pm$ 0.70 & 8.65 $\pm$ 1.58 & 8.60 $\pm$ 0.86 & 8.92 $\pm$ 1.34 & 9.26 $\pm$ 1.38 \\ 
\multicolumn{1}{c|}{} & \multicolumn{1}{l|}{\textcolor{black}{SENet}} & \multicolumn{1}{l|}{\textcolor{black}{LFCC + MPE}} & \textcolor{black}{39.00 $\pm$ 0.67} & \textcolor{black}{38.38 $\pm$ 0.50} & \textcolor{black}{36.11 $\pm$ 0.30} & \textcolor{black}{35.19 $\pm$ 0.02} & \textcolor{black}{35.69 $\pm$ 1.13} & \textcolor{black}{35.26 $\pm$ 0.95} \\ 
\multicolumn{1}{c|}{} & \multicolumn{1}{l|}{\textcolor{black}{FFN-WavLM}} & \multicolumn{1}{l|}{\textcolor{black}{Raw Waveform}} & \textcolor{black}{11.02 $\pm$ 2.90} & \textcolor{black}{11.29 $\pm$ 2.90} & \textcolor{black}{10.68 $\pm$ 3.79} & \textcolor{black}{10.11 $\pm$ 2.59} & \textcolor{black}{8.67 $\pm$ 3.44} & \textcolor{black}{8.58 $\pm$ 2.16} \\ \hline
\multicolumn{1}{c|}{\multirow{6}{*}{\begin{tabular}[c]{@{}c@{}}Audio\\ Contrastive/\\ Non-\\ contrastive\end{tabular}}} & \multicolumn{1}{l|}{SSL-ACL} & \multicolumn{1}{l|}{Log-mel} & 35.77 $\pm$ 1.57 & 31.94 $\pm$ 0.56 & 27.42 $\pm$ 1.02 & 26.36 $\pm$ 1.02 & 26.05 $\pm$ 0.38 & 26.00 $\pm$ 0.53 \\
\multicolumn{1}{c|}{} & \multicolumn{1}{l|}{BYOL-A} & \multicolumn{1}{l|}{Log-mel} & 23.55 $\pm$ 0.63 & 22.08 $\pm$ 0.74 & 22.92 $\pm$ 0.18 & 21.69 $\pm$ 0.79 & 21.91 $\pm$ 0.81 & 21.77 $\pm$ 0.13 \\
\multicolumn{1}{c|}{} & \multicolumn{1}{l|}{COLA} & \multicolumn{1}{l|}{Log-mel} & 27.25 $\pm$ 1.40 & 25.49 $\pm$ 1.70 & 23.85 $\pm$ 1.02 & 23.29 $\pm$ 0.16 & 23.17 $\pm$ 0.23 & 23.93 $\pm$ 0.39 \\
\multicolumn{1}{c|}{} & \multicolumn{1}{l|}{MoCo} & \multicolumn{1}{l|}{MFCC} & 27.89 $\pm$ 1.15 & 28.35 $\pm$ 0.84 & 28.90 $\pm$ 0.76 & 28.45 $\pm$ 1.57 & 28.08 $\pm$ 0.40 & 26.99 $\pm$ 1.12 \\
\multicolumn{1}{c|}{} & \multicolumn{1}{l|}{SimCLR} & \multicolumn{1}{l|}{Log-mel} & 24.95 $\pm$ 1.18 & 24.72 $\pm$ 1.13 & 22.64 $\pm$ 0.62 & 20.92 $\pm$ 1.60 & 20.92 $\pm$ 0.77 & 20.47 $\pm$ 0.70 \\
\multicolumn{1}{c|}{} & \multicolumn{1}{l|}{VG-NCL-$\tau$} & \multicolumn{1}{l|}{Wav2Vec2} & 14.30 $\pm$ 1.07 & 10.03 $\pm$ 1.79 & 8.75 $\pm$ 1.07 & 8.82 $\pm$ 0.34 & 9.56 $\pm$ 0.28 & 9.06 $\pm$ 0.39 \\
\multicolumn{1}{c|}{} & \multicolumn{1}{l|}{VG-NCL-$\varsigma$} & \multicolumn{1}{l|}{Wav2Vec2} & 9.35 $\pm$ 0.83 & 12.67 $\pm$ 5.10 & 7.77 $\pm$ 0.62 & 8.76 $\pm$ 1.48 & 8.08 $\pm$ 0.50 & 10.57 $\pm$ 3.98 \\
\multicolumn{1}{c|}{} & \multicolumn{1}{l|}{\textbf{SIGNL}} & \multicolumn{1}{l|}{Wav2Vec2} & \textbf{7.88 $\pm$ 2.11} & \textbf{6.76 $\pm$ 1.05} & \textbf{6.85 $\pm$ 0.36} & \textbf{7.01 $\pm$ 1.40} & \textbf{6.51 $\pm$ 0.52} & \textbf{7.21 $\pm$ 0.61} \\ \hline
 &  &  &  &  &  &  &  &  \\ \hline
\multicolumn{1}{c|}{\multirow{2}{*}{\textbf{Categories}}} & \multicolumn{1}{c|}{\multirow{2}{*}{\textbf{Methods}}} & \multicolumn{1}{c|}{\multirow{2}{*}{\textbf{Front-end}}} & \multicolumn{6}{c}{\textbf{ASVspoof 5}} \\ \cline{4-9} 
\multicolumn{1}{c|}{} & \multicolumn{1}{c|}{} & \multicolumn{1}{c|}{} & \textbf{5\%} & \textbf{10\%} & \textbf{30\%} & \textbf{50\%} & \textbf{80\%} & \textbf{Full} \\ \hline
\multicolumn{1}{c|}{\multirow{6}{*}{Supervised}} & \multicolumn{1}{l|}{LCNN} & \multicolumn{1}{l|}{LFCC} & 25.06 $\pm$ 1.88 & 22.07 $\pm$ 3.53 & 19.40 $\pm$ 0.32 & 17.29 $\pm$ 0.23 & 17.25 $\pm$ 1.08 & 17.65 $\pm$ 1.18 \\
\multicolumn{1}{c|}{} & \multicolumn{1}{l|}{LCNN} & \multicolumn{1}{l|}{Whisper} & 20.70 $\pm$ 0.58 & 19.08 $\pm$ 0.10 & 16.82 $\pm$ 5.02 & 19.37 $\pm$ 0.14 & 17.61 $\pm$ 1.40 & 18.86 $\pm$ 1.94 \\
\multicolumn{1}{c|}{} & \multicolumn{1}{l|}{LCNN} & \multicolumn{1}{l|}{Wav2Vec2} & 15.89 $\pm$ 4.61 & 11.77 $\pm$ 4.31 & 5.74 $\pm$ 2.91 & 4.15 $\pm$ 2.14 & 3.85 $\pm$ 0.91 & 5.92 $\pm$ 0.69 \\
\multicolumn{1}{c|}{} & \multicolumn{1}{l|}{AASIST} & \multicolumn{1}{l|}{Raw Waveform} & 31.41 $\pm$ 8.78 & 22.96 $\pm$ 0.10 & 25.29 $\pm$ 3.44 & 20.21 $\pm$ 1.29 & 21.48 $\pm$ 3.01 & 22.37 $\pm$ 0.95 \\
\multicolumn{1}{c|}{} & \multicolumn{1}{l|}{AASIST} & \multicolumn{1}{l|}{Whisper} & 21.91 $\pm$ 6.84 & 18.60 $\pm$ 6.23 & 19.05 $\pm$ 5.92 & 14.31 $\pm$ 1.27 & 11.81 $\pm$ 2.78 & 16.32 $\pm$ 4.92 \\
\multicolumn{1}{c|}{} & \multicolumn{1}{l|}{AASIST} & \multicolumn{1}{l|}{Wav2Vec2} & 6.83 $\pm$ 2.68 & 6.91 $\pm$ 1.99 & 4.91 $\pm$ 2.93 & 10.34 $\pm$ 2.63 & 3.05 $\pm$ 1.17 & 3.00 $\pm$ 0.76 \\ 
\multicolumn{1}{c|}{} & \multicolumn{1}{l|}{\textcolor{black}{SENet}} & \multicolumn{1}{l|}{\textcolor{black}{LFCC + MPE}} & \textcolor{black}{18.89 $\pm$ 2.70} & \textcolor{black}{16.57 $\pm$ 3.56} & \textcolor{black}{16.13 $\pm$ 3.05} & \textcolor{black}{15.10 $\pm$ 2.86} & \textcolor{black}{16.15 $\pm$ 2.87} & \textcolor{black}{16.35 $\pm$ 2.52}\\ 
\multicolumn{1}{c|}{} & \multicolumn{1}{l|}{\textcolor{black}{FFN-WavLM}} & \multicolumn{1}{l|}{\textcolor{black}{Raw Waveform}} & \textcolor{black}{34.81 $\pm$ 7.14} & \textcolor{black}{31.14 $\pm$ 9.40} & \textcolor{black}{29.77 $\pm$ 10.15} & \textcolor{black}{23.70 $\pm$ 11.04} & \textcolor{black}{24.38 $\pm$ 11.95} & \textcolor{black}{21.48 $\pm$ 11.17}\\ \hline
\multicolumn{1}{c|}{\multirow{6}{*}{\begin{tabular}[c]{@{}c@{}}Audio\\ Contrastive/\\ Non-\\ contrastive\end{tabular}}} & \multicolumn{1}{l|}{SSL-ACL} & \multicolumn{1}{l|}{Log-mel} & 18.70 $\pm$ 0.33 & 17.15 $\pm$ 0.15 & 15.24 $\pm$ 0.13 & 15.00 $\pm$ 0.48 & 14.20 $\pm$ 0.37 & 13.95 $\pm$ 1.13 \\
\multicolumn{1}{c|}{} & \multicolumn{1}{l|}{BYOL-A} & \multicolumn{1}{l|}{Log-mel} & 16.19 $\pm$ 1.64 & 18.72 $\pm$ 1.97 & 14.94 $\pm$ 2.89 & 16.75 $\pm$ 2.79 & 14.57 $\pm$ 2.06 & 13.40 $\pm$ 2.25 \\
\multicolumn{1}{c|}{} & \multicolumn{1}{l|}{COLA} & \multicolumn{1}{l|}{Log-mel} & 14.71 $\pm$ 1.16 & 14.63 $\pm$ 1.14 & 14.17 $\pm$ 0.21 & 14.47 $\pm$ 0.82 & 13.52 $\pm$ 0.84 & 13.43 $\pm$ 1.07 \\
\multicolumn{1}{c|}{} & \multicolumn{1}{l|}{MoCo} & \multicolumn{1}{l|}{MFCC} & 30.59 $\pm$ 0.82 & 29.94 $\pm$ 0.56 & 26.59 $\pm$ 1.88 & 25.65 $\pm$ 0.99 & 24.93 $\pm$ 0.38 & 24.53 $\pm$ 1.78 \\
\multicolumn{1}{c|}{} & \multicolumn{1}{l|}{SimCLR} & \multicolumn{1}{l|}{Log-mel} & 17.51 $\pm$ 1.03 & 15.22 $\pm$ 1.58 & 16.29 $\pm$ 1.83 & 15.71 $\pm$ 2.14 & 13.49 $\pm$ 0.72 & 12.78 $\pm$ 1.93 \\
\multicolumn{1}{c|}{} & \multicolumn{1}{l|}{VG-NCL-$\tau$} & \multicolumn{1}{l|}{Wav2Vec2} & 4.84 $\pm$ 2.21 & 6.16 $\pm$ 0.82 & 6.70 $\pm$ 2.12 & 6.05 $\pm$ 2.06 & 5.17 $\pm$ 0.90 & 4.33 $\pm$ 1.97 \\
\multicolumn{1}{c|}{} & \multicolumn{1}{l|}{VG-NCL-$\varsigma$} & \multicolumn{1}{l|}{Wav2Vec2} & 4.71 $\pm$ 1.91 & 6.71 $\pm$ 2.33 & 5.32 $\pm$ 1.20 & 9.67 $\pm$ 5.16 & 6.36 $\pm$ 2.25 & 6.25 $\pm$ 1.15 \\
\multicolumn{1}{c|}{} & \multicolumn{1}{l|}{\textbf{SIGNL}} & \multicolumn{1}{l|}{Wav2Vec2} & \textbf{3.95 $\pm$ 1.34} & \textbf{2.43 $\pm$ 0.88} & \textbf{2.86 $\pm$ 1.30} & \textbf{2.65 $\pm$ 1.21} & \textbf{1.89 $\pm$ 0.97} & \textbf{2.33 $\pm$ 0.66} \\ \hline
 &  &  &  &  &  &  &  &  \\ \hline
\multicolumn{1}{c|}{\multirow{2}{*}{\textbf{Categories}}} & \multicolumn{1}{c|}{\multirow{2}{*}{\textbf{Methods}}} & \multicolumn{1}{c|}{\multirow{2}{*}{\textbf{Front-end}}} & \multicolumn{6}{c}{\textbf{CFAD}} \\ \cline{4-9} 
\multicolumn{1}{c|}{} & \multicolumn{1}{c|}{} & \multicolumn{1}{c|}{} & \textbf{5\%} & \textbf{10\%} & \textbf{30\%} & \textbf{50\%} & \textbf{80\%} & \textbf{Full} \\ \hline
\multicolumn{1}{c|}{\multirow{6}{*}{Supervised}} & \multicolumn{1}{l|}{LCNN} & \multicolumn{1}{l|}{LFCC} & 25.42 $\pm$ 0.50 & 24.94 $\pm$ 1.82 & 22.14 $\pm$ 2.44 & 22.36 $\pm$ 0.76 & 20.87 $\pm$ 0.54 & 20.02 $\pm$ 2.18 \\
\multicolumn{1}{c|}{} & \multicolumn{1}{l|}{LCNN} & \multicolumn{1}{l|}{Whisper} & 16.58 $\pm$ 2.08 & 16.73 $\pm$ 2.23 & 13.19 $\pm$ 0.48 & 12.46 $\pm$ 0.77 & 11.32 $\pm$ 0.66 & 10.52 $\pm$ 0.50 \\
\multicolumn{1}{c|}{} & \multicolumn{1}{l|}{LCNN} & \multicolumn{1}{l|}{Wav2Vec2} & 10.38 $\pm$ 0.08 & 11.19 $\pm$ 1.81 & 11.33 $\pm$ 1.26 & 8.71 $\pm$ 0.39 & 9.64 $\pm$ 0.01 & 9.09 $\pm$ 0.55 \\
\multicolumn{1}{c|}{} & \multicolumn{1}{l|}{AASIST} & \multicolumn{1}{l|}{Raw Waveform} & 29.55 $\pm$ 2.06 & 23.45 $\pm$ 1.69 & 18.33 $\pm$ 0.83 & 17.62 $\pm$ 0.34 & 17.76 $\pm$ 0.87 & 17.04 $\pm$ 0.21 \\
\multicolumn{1}{c|}{} & \multicolumn{1}{l|}{AASIST} & \multicolumn{1}{l|}{Whisper} & 16.72 $\pm$ 2.60 & 14.99 $\pm$ 1.29 & 13.39 $\pm$ 0.54 & 12.28 $\pm$ 0.17 & 12.74 $\pm$ 1.10 & 12.47 $\pm$ 0.93 \\
\multicolumn{1}{c|}{} & \multicolumn{1}{l|}{AASIST} & \multicolumn{1}{l|}{Wav2Vec2} & 11.96 $\pm$ 3.22 & 12.04 $\pm$ 3.08 & 9.80 $\pm$ 0.58 & 9.76 $\pm$ 1.20 & 9.11 $\pm$ 0.40 & 8.83 $\pm$ 0.52 \\ 
\multicolumn{1}{c|}{} & \multicolumn{1}{l|}{\textcolor{black}{SENet}} & \multicolumn{1}{l|}{\textcolor{black}{LFCC + MPE}} & \textcolor{black}{27.16 $\pm$ 0.68} & \textcolor{black}{25.18 $\pm$ 0.78} & \textcolor{black}{24.43 $\pm$ 1.08} & \textcolor{black}{23.36 $\pm$ 0.72} & \textcolor{black}{23.85 $\pm$ 1.31} & \textcolor{black}{23.80 $\pm$ 1.10} \\ 
\multicolumn{1}{c|}{} & \multicolumn{1}{l|}{\textcolor{black}{FFN-WavLM}} & \multicolumn{1}{l|}{\textcolor{black}{Raw Waveform}} & \textcolor{black}{10.69 $\pm$ 2.69} & \textcolor{black}{10.38 $\pm$ 1.39} & \textcolor{black}{8.90 $\pm$ 0.26} & \textcolor{black}{8.58 $\pm$ 0.16} & \textcolor{black}{8.49 $\pm$ 0.17} & \textcolor{black}{\textbf{8.41 $\pm$ 0.12}}\\ \hline
\multicolumn{1}{c|}{\multirow{6}{*}{\begin{tabular}[c]{@{}c@{}}Audio\\ Contrastive/\\ Non-\\ contrastive\end{tabular}}} & \multicolumn{1}{l|}{SSL-ACL} & \multicolumn{1}{l|}{Log-mel} & 29.44 $\pm$ 0.37 & 36.20 $\pm$ 0.58 & 33.47 $\pm$ 0.47 & 31.59 $\pm$ 0.32 & 30.67 $\pm$ 0.24 & 29.14 $\pm$ 0.85 \\
\multicolumn{1}{c|}{} & \multicolumn{1}{l|}{BYOL-A} & \multicolumn{1}{l|}{Log-mel} & 21.94 $\pm$ 2.10 & 20.32 $\pm$ 3.17 & 16.81 $\pm$ 1.60 & 16.19 $\pm$ 1.09 & 16.68 $\pm$ 0.43 & 15.84 $\pm$ 0.90 \\
\multicolumn{1}{c|}{} & \multicolumn{1}{l|}{COLA} & \multicolumn{1}{l|}{Log-mel} & 21.14 $\pm$ 0.10 & 21.48 $\pm$ 0.75 & 18.44 $\pm$ 0.62 & 18.70 $\pm$ 0.20 & 16.45 $\pm$ 1.46 & 18.52 $\pm$ 1.30 \\
\multicolumn{1}{c|}{} & \multicolumn{1}{l|}{MoCo} & \multicolumn{1}{l|}{MFCC} & 28.59 $\pm$ 1.07 & 24.00 $\pm$ 1.60 & 22.92 $\pm$ 1.59 & 20.96 $\pm$ 1.46 & 20.39 $\pm$ 1.55 & 20.10 $\pm$ 0.61 \\
\multicolumn{1}{c|}{} & \multicolumn{1}{l|}{SimCLR} & \multicolumn{1}{l|}{Log-mel} & 27.23 $\pm$ 1.01 & 20.94 $\pm$ 1.11 & 20.50 $\pm$ 0.49 & 17.56 $\pm$ 1.02 & 17.59 $\pm$ 0.81 & 16.87 $\pm$ 0.36 \\
\multicolumn{1}{c|}{} & \multicolumn{1}{l|}{VG-NCL-$\tau$} & \multicolumn{1}{l|}{Wav2Vec2} & \textbf{9.48 $\pm$ 0.40} & 9.31 $\pm$ 0.53 & \textbf{8.30 $\pm$ 0.46} & 9.24 $\pm$ 0.88 & 8.94 $\pm$ 0.45 & 8.72 $\pm$ 0.10 \\
\multicolumn{1}{c|}{} & \multicolumn{1}{l|}{VG-NCL-$\varsigma$} & \multicolumn{1}{l|}{Wav2Vec2} & 10.07 $\pm$ 1.05 & 11.75 $\pm$ 1.45 & 8.92 $\pm$ 0.17 & 9.12 $\pm$ 0.94 & 8.98 $\pm$ 0.56 & 9.48 $\pm$ 0.49 \\
\multicolumn{1}{c|}{} & \multicolumn{1}{l|}{\textbf{SIGNL}} & \multicolumn{1}{l|}{Wav2Vec2} & 9.90 $\pm$ 0.20 & \textbf{8.84 $\pm$ 0.10} & 8.77 $\pm$ 0.20 & \textbf{8.46 $\pm$ 0.21} & \textbf{8.83 $\pm$ 0.50} & 8.44 $\pm$ 0.12 \\ \hline
\end{tabular}
}
\end{table*}

\begin{table*}[htbp]
\centering
\caption{EER (\%) comparison across different levels of label information for cross-domain evaluation on In-The-Wild dataset. \textcolor{black}{All values of the methods\protect\footnotemark[2]} represent mean $\pm$ standard deviation. Lower values: better performance ($\downarrow$). \textbf{Bold}: best result.}
\label{tab:contrast_itw}
\resizebox{\textwidth}{!}{%
\begin{tabular}{cllcccccc}
\hline
\multicolumn{1}{c|}{\multirow{2}{*}{\textbf{Categories}}} & \multicolumn{1}{c|}{\multirow{2}{*}{\textbf{Methods}}} & \multicolumn{1}{l|}{\multirow{2}{*}{\textbf{Front-end}}} & \multicolumn{6}{c}{\textbf{Trained on ASVspoof 2021 DF, Tested on In-The-Wild}} \\ \cline{4-9} 
\multicolumn{1}{c|}{} & \multicolumn{1}{c|}{} & \multicolumn{1}{l|}{} & \textbf{5\%} & \textbf{10\%} & \textbf{30\%} & \textbf{50\%} & \textbf{80\%} & \textbf{Full} \\ \hline
\multicolumn{1}{c|}{\multirow{6}{*}{Supervised}} & \multicolumn{1}{l|}{LCNN} & \multicolumn{1}{l|}{LFCC} & 51.04 $\pm$ 5.33 & 48.36 $\pm$ 1.47 & 50.09 $\pm$ 2.88 & 49.67 $\pm$ 0.80 & 49.37 $\pm$ 3.66 & 47.90 $\pm$ 1.48 \\
\multicolumn{1}{c|}{} & \multicolumn{1}{l|}{LCNN} & \multicolumn{1}{l|}{Whisper} & 33.78 $\pm$ 2.67 & 29.19 $\pm$ 5.36 & 26.98 $\pm$ 2.76 & 30.03 $\pm$ 4.06 & 27.97 $\pm$ 2.43 & 28.56 $\pm$ 5.18 \\
\multicolumn{1}{c|}{} & \multicolumn{1}{l|}{LCNN} & \multicolumn{1}{l|}{Wav2Vec2} & 39.44 $\pm$ 1.54 & 32.01 $\pm$ 9.96 & 33.16 $\pm$ 0.33 & 27.12 $\pm$ 2.03 & 18.85 $\pm$ 0.75 & 14.43 $\pm$ 5.53  \\
\multicolumn{1}{c|}{} & \multicolumn{1}{l|}{AASIST} & \multicolumn{1}{l|}{Raw Waveform} & 45.92 $\pm$ 3.31 & 41.91 $\pm$ 4.90 & 45.10 $\pm$ 6.46 & 45.96 $\pm$ 5.36 & 47.33 $\pm$ 3.48 & 48.57 $\pm$ 3.90 \\
\multicolumn{1}{c|}{} & \multicolumn{1}{l|}{AASIST} & \multicolumn{1}{l|}{Whisper} & 30.45 $\pm$ 1.40 & 29.19 $\pm$ 0.49 & 28.19 $\pm$ 2.06 & 28.85 $\pm$ 1.00 & 27.64 $\pm$ 1.80 & 28.29 $\pm$ 3.27 \\
\multicolumn{1}{c|}{} & \multicolumn{1}{l|}{AASIST} & \multicolumn{1}{l|}{Wav2Vec2} & 36.16 $\pm$ 6.08 & 18.71 $\pm$ 4.48 & 13.96 $\pm$ 1.79 & 12.20 $\pm$ 2.82 & 16.06 $\pm$ 3.94 & 12.68 $\pm$ 3.50 \\ 
\multicolumn{1}{c|}{} & \multicolumn{1}{l|}{\textcolor{black}{SENet}} & \multicolumn{1}{l|}{\textcolor{black}{LFCC + MPE}} & \textcolor{black}{33.74 $\pm$ 3.79} & \textcolor{black}{41.44 $\pm$ 3.58} & \textcolor{black}{53.03 $\pm$ 4.78} & \textcolor{black}{54.24 $\pm$ 6.06} & \textcolor{black}{53.04 $\pm$ 3.69} & \textcolor{black}{51.70 $\pm$ 3.98} \\ 
\multicolumn{1}{c|}{} & \multicolumn{1}{l|}{\textcolor{black}{FFN-WavLM}} & \multicolumn{1}{l|}{\textcolor{black}{Raw Waveform}} & \textcolor{black}{21.10 $\pm$ 6.10} & \textcolor{black}{27.66 $\pm$ 3.33} & \textcolor{black}{25.80 $\pm$ 4.92} & \textcolor{black}{21.70 $\pm$ 2.51} & \textcolor{black}{19.84 $\pm$ 4.99} & \textcolor{black}{16.86 $\pm$ 5.71} \\ \hline
\multicolumn{1}{c|}{\multirow{6}{*}{\begin{tabular}[c]{@{}c@{}}Audio\\ Contrastive/\\ Non-\\ contrastive\end{tabular}}} & \multicolumn{1}{l|}{SSL-ACL} & \multicolumn{1}{l|}{Log-mel} & 53.13 $\pm$ 1.78 & 57.03 $\pm$ 1.56 & 66.37 $\pm$ 6.15 & 73.85 $\pm$ 1.61 & 75.58 $\pm$ 4.56 & 76.89 $\pm$ 0.99 \\
\multicolumn{1}{c|}{} & \multicolumn{1}{l|}{BYOL-A} & \multicolumn{1}{l|}{Log-mel} & 59.75 $\pm$ 5.87 & 63.18 $\pm$ 6.60 & 66.24 $\pm$ 1.72 & 69.41 $\pm$ 7.12 & 69.63 $\pm$ 2.97 & 70.47 $\pm$ 3.81 \\
\multicolumn{1}{c|}{} & \multicolumn{1}{l|}{COLA} & \multicolumn{1}{l|}{Log-mel} & 56.42 $\pm$ 8.77 & 65.34 $\pm$ 3.93 & 74.62 $\pm$ 5.24 & 78.89 $\pm$ 2.31 & 77.00 $\pm$ 5.39 & 76.61 $\pm$ 0.14 \\
\multicolumn{1}{c|}{} & \multicolumn{1}{l|}{MoCo} & \multicolumn{1}{l|}{MFCC} & 59.91 $\pm$ 4.84 & 57.62 $\pm$ 8.77 & 55.09 $\pm$ 4.29 & 53.55 $\pm$ 5.49 & 64.43 $\pm$ 0.83 & 52.43 $\pm$ 0.33 \\
\multicolumn{1}{c|}{} & \multicolumn{1}{l|}{SimCLR} & \multicolumn{1}{l|}{Log-mel} & 38.15 $\pm$ 4.30 & 41.52 $\pm$ 3.37 & 36.09 $\pm$ 5.27 & 33.36 $\pm$ 7.47 & 28.79 $\pm$ 1.37 & 30.08 $\pm$ 2.49 \\
\multicolumn{1}{c|}{} & \multicolumn{1}{l|}{VG-NCL-$\tau$} & \multicolumn{1}{l|}{Wav2Vec2} & 25.02 $\pm$ 5.69 & 19.09 $\pm$ 3.43 & 16.47 $\pm$ 3.21 & 17.30 $\pm$ 2.74 & 17.14 $\pm$ 1.81 & 15.59 $\pm$ 1.21 \\
\multicolumn{1}{c|}{} & \multicolumn{1}{l|}{VG-NCL-$\varsigma$} & \multicolumn{1}{l|}{Wav2Vec2} &\textbf{14.45 $\pm$ 3.22} & 19.15 $\pm$ 4.30 & 11.61 $\pm$ 2.70 & 13.37 $\pm$ 2.04 & 15.16 $\pm$ 1.64 & 10.80 $\pm$ 0.65 \\
\multicolumn{1}{c|}{} & \multicolumn{1}{l|}{\textbf{SIGNL}} & \multicolumn{1}{l|}{Wav2Vec2} & 14.85 $\pm$ 2.21 & \textbf{12.51 $\pm$ 0.50} & \textbf{10.18 $\pm$ 1.43} & \textbf{9.67 $\pm$ 1.55} & \textbf{11.18 $\pm$ 1.50} & \textbf{10.89 $\pm$ 3.10} \\ \hline
 &  &  &  &  &  &  &  &  \\ \hline
\multicolumn{1}{c|}{\multirow{2}{*}{\textbf{Categories}}} & \multicolumn{1}{c|}{\multirow{2}{*}{\textbf{Methods}}} & \multicolumn{1}{c|}{\multirow{2}{*}{\textbf{Front-end}}} & \multicolumn{6}{c}{\textbf{Trained on ASVspoof 5, Tested on In-The-Wild}} \\ \cline{4-9} 
\multicolumn{1}{c|}{} & \multicolumn{1}{c|}{} & \multicolumn{1}{c|}{} & \textbf{5\%} & \textbf{10\%} & \textbf{30\%} & \textbf{50\%} & \textbf{80\%} & \textbf{Full} \\ \hline
\multicolumn{1}{c|}{\multirow{6}{*}{Supervised}} & \multicolumn{1}{l|}{LCNN} & \multicolumn{1}{l|}{LFCC} & 46.26 $\pm$ 1.21 & 49.58 $\pm$ 1.38 & 52.79 $\pm$ 1.77 & 51.68 $\pm$ 2.46 & 51.89 $\pm$ 3.08 & 53.46 $\pm$ 1.17 \\
\multicolumn{1}{c|}{} & \multicolumn{1}{l|}{LCNN} & \multicolumn{1}{l|}{Whisper} & 33.78 $\pm$ 3.28 & 29.19 $\pm$ 6.56 & 26.98 $\pm$ 3.38 & 30.03 $\pm$ 4.97 & 27.97 $\pm$ 2.97 & 28.56 $\pm$ 6.34 \\
\multicolumn{1}{c|}{} & \multicolumn{1}{l|}{LCNN} & \multicolumn{1}{l|}{Wav2Vec2} & 37.33 $\pm$ 4.10 & 42.64 $\pm$ 3.67 & 30.76 $\pm$ 1.07 & 36.21 $\pm$ 5.97 & 29.16 $\pm$ 5.02 & 32.71 $\pm$ 5.02 \\
\multicolumn{1}{c|}{} & \multicolumn{1}{l|}{AASIST} & \multicolumn{1}{l|}{Raw Waveform} & 34.93 $\pm$ 3.91 & 29.51 $\pm$ 2.06 & \textbf{26.91 $\pm$ 0.59 }& 26.73 $\pm$ 1.46 & 25.56 $\pm$ 2.03 & 26.86 $\pm$ 2.10 \\
\multicolumn{1}{c|}{} & \multicolumn{1}{l|}{AASIST} & \multicolumn{1}{l|}{Whisper} & 32.91 $\pm$ 7.97 & 31.05 $\pm$ 3.91 & 29.62 $\pm$ 2.41 & 26.54 $\pm$ 2.06 & 24.86 $\pm$ 3.99 & \textbf{24.59 $\pm$ 3.19} \\
\multicolumn{1}{c|}{} & \multicolumn{1}{l|}{AASIST} & \multicolumn{1}{l|}{Wav2Vec2} & 37.48 $\pm$ 2.06 & 39.29 $\pm$ 7.22 & 36.11 $\pm$ 6.92 & 46.32 $\pm$ 3.76 & 28.10 $\pm$ 1.94 & 31.88 $\pm$ 3.84 \\
\multicolumn{1}{c|}{} & \multicolumn{1}{l|}{\textcolor{black}{SENet}} & \multicolumn{1}{l|}{\textcolor{black}{LFCC + MPE}} & \textcolor{black}{59.02 $\pm$ 3.00} & \textcolor{black}{56.46 $\pm$ 2.92} & \textcolor{black}{54.75 $\pm$ 0.33} & \textcolor{black}{53.90 $\pm$ 2.09} & \textcolor{black}{54.31 $\pm$ 2.41} & \textcolor{black}{53.65 $\pm$ 1.13} \\
\multicolumn{1}{c|}{} & \multicolumn{1}{l|}{\textcolor{black}{FFN-WavLM}} & \multicolumn{1}{l|}{\textcolor{black}{Raw Waveform}} & \textcolor{black}{35.03 $\pm$ 7.28} & \textcolor{black}{35.44 $\pm$ 12.23} & \textcolor{black}{32.85 $\pm$ 7.41} & \textcolor{black}{27.06 $\pm$ 6.83} & \textcolor{black}{31.33 $\pm$ 8.29} & \textcolor{black}{28.43 $\pm$ 5.62}\\ \hline
\multicolumn{1}{c|}{\multirow{6}{*}{\begin{tabular}[c]{@{}c@{}}Audio\\ Contrastive/\\ Non-\\ contrastive\end{tabular}}} & \multicolumn{1}{l|}{SSL-ACL} & \multicolumn{1}{l|}{Log-mel} & 78.87 $\pm$ 1.27 & 77.71 $\pm$ 0.12 & 71.55 $\pm$ 0.59 & 71.18 $\pm$ 1.66 & 68.42 $\pm$ 2.40 & 71.43 $\pm$ 3.41 \\
\multicolumn{1}{c|}{} & \multicolumn{1}{l|}{BYOL-A} & \multicolumn{1}{l|}{Log-mel} & 79.27 $\pm$ 2.91 & 77.12 $\pm$ 5.02 & 76.73 $\pm$ 3.03 & 73.73 $\pm$ 3.28 & 75.25 $\pm$ 1.97 & 73.72 $\pm$ 1.65 \\
\multicolumn{1}{c|}{} & \multicolumn{1}{l|}{COLA} & \multicolumn{1}{l|}{Log-mel} & 68.57 $\pm$ 2.56 & 75.95 $\pm$ 7.81 & 75.24 $\pm$ 3.93 & 78.22 $\pm$ 1.32 & 76.35 $\pm$ 4.00 & 71.74 $\pm$ 2.64 \\
\multicolumn{1}{c|}{} & \multicolumn{1}{l|}{MoCo} & \multicolumn{1}{l|}{MFCC} & 55.02 $\pm$ 3.07 & 65.81 $\pm$ 3.12 & 54.78 $\pm$ 2.79 & 74.98 $\pm$ 4.36 & 69.11 $\pm$ 0.10 & 69.56 $\pm$ 4.11 \\
\multicolumn{1}{c|}{} & \multicolumn{1}{l|}{SimCLR} & \multicolumn{1}{l|}{Log-mel} & 39.96 $\pm$ 2.52 & 45.97 $\pm$ 5.74 & 45.62 $\pm$ 5.79 & 39.46 $\pm$ 0.51 & 56.59 $\pm$ 1.50 & 52.24 $\pm$ 2.46 \\
\multicolumn{1}{c|}{} & \multicolumn{1}{l|}{VG-NCL-$\tau$} & \multicolumn{1}{l|}{Wav2Vec2} & 40.35 $\pm$ 3.10 & 44.24 $\pm$ 0.57 & 41.53 $\pm$ 4.17 & 42.57 $\pm$ 11.76 & 38.71 $\pm$ 4.40 & 34.51 $\pm$ 7.90 \\
\multicolumn{1}{c|}{} & \multicolumn{1}{l|}{VG-NCL-$\varsigma$} & \multicolumn{1}{l|}{Wav2Vec2} & 37.67 $\pm$ 0.33 & 39.49 $\pm$ 7.49 & 32.10 $\pm$ 2.48 & 33.49 $\pm$ 10.18 & 29.08 $\pm$ 6.17 & 27.46 $\pm$ 2.84 \\
\multicolumn{1}{c|}{} & \multicolumn{1}{l|}{\textbf{SIGNL}} & \multicolumn{1}{l|}{Wav2Vec2} & \textbf{31.72 $\pm$ 2.59} & \textbf{27.62 $\pm$ 4.46} & 28.66 $\pm$ 3.57 & \textbf{24.24 $\pm$ 5.15} & \textbf{22.99 $\pm$ 2.41} & 27.43 $\pm$ 2.29 \\ \hline
 &  &  &  &  &  &  &  &  \\ \hline
\multicolumn{1}{c|}{\multirow{2}{*}{\textbf{Categories}}} & \multicolumn{1}{c|}{\multirow{2}{*}{\textbf{Methods}}} & \multicolumn{1}{c|}{\multirow{2}{*}{\textbf{Front-end}}} & \multicolumn{6}{c}{\textbf{Trained on CFAD, Tested on In-The-Wild}} \\ \cline{4-9} 
\multicolumn{1}{c|}{} & \multicolumn{1}{c|}{} & \multicolumn{1}{c|}{} & \textbf{5\%} & \textbf{10\%} & \textbf{30\%} & \textbf{50\%} & \textbf{80\%} & \textbf{Full} \\ \hline
\multicolumn{1}{c|}{\multirow{6}{*}{Supervised}} & \multicolumn{1}{l|}{LCNN} & \multicolumn{1}{l|}{LFCC} & 48.39 $\pm$ 2.61 & 53.69 $\pm$ 3.08 & 56.51 $\pm$ 1.11 & 58.37 $\pm$ 1.71 & 60.02 $\pm$ 2.89 & 56.83 $\pm$ 2.44 \\
\multicolumn{1}{c|}{} & \multicolumn{1}{l|}{LCNN} & \multicolumn{1}{l|}{Whisper} & 47.61 $\pm$ 2.80 & 47.89 $\pm$ 2.91 & 43.64 $\pm$ 2.16 & 46.29 $\pm$ 2.70 & 47.92 $\pm$ 0.44 & 48.73 $\pm$ 1.26 \\
\multicolumn{1}{c|}{} & \multicolumn{1}{l|}{LCNN} & \multicolumn{1}{l|}{Wav2Vec2} & 13.04 $\pm$ 2.90 & 15.03 $\pm$ 0.80 & 13.27 $\pm$ 4.01 & \textbf{8.16 $\pm$ 0.16} & 8.91 $\pm$ 0.74 & 8.35 $\pm$ 2.05 \\
\multicolumn{1}{c|}{} & \multicolumn{1}{l|}{AASIST} & \multicolumn{1}{l|}{Raw Waveform} & 56.12 $\pm$ 6.95 & 57.16 $\pm$ 1.38 & 57.06 $\pm$ 6.31 & 62.22 $\pm$ 0.28 & 56.89 $\pm$ 5.43 & 60.60 $\pm$ 4.39 \\
\multicolumn{1}{c|}{} & \multicolumn{1}{l|}{AASIST} & \multicolumn{1}{l|}{Whisper} & 44.03 $\pm$ 6.48 & 47.66 $\pm$ 3.76 & 48.58 $\pm$ 2.21 & 50.08 $\pm$ 3.46 & 49.80 $\pm$ 2.96 & 48.66 $\pm$ 0.65 \\
\multicolumn{1}{c|}{} & \multicolumn{1}{l|}{AASIST} & \multicolumn{1}{l|}{Wav2Vec2} & 13.74 $\pm$ 1.85 & 13.06 $\pm$ 3.05 & \textbf{9.74 $\pm$ 2.56} & 10.23 $\pm$ 0.40 & 10.12 $\pm$ 3.52 & 11.35 $\pm$ 0.93 \\ 
\multicolumn{1}{c|}{} & \multicolumn{1}{l|}{\textcolor{black}{SENet}} & \multicolumn{1}{l|}{\textcolor{black}{LFCC + MPE}} & \textcolor{black}{70.38 $\pm$ 5.06} & \textcolor{black}{73.54 $\pm$ 2.30} & \textcolor{black}{77.33 $\pm$ 2.48} & \textcolor{black}{79.50 $\pm$ 2.60} & \textcolor{black}{79.93 $\pm$ 3.73} & \textcolor{black}{79.48 $\pm$ 2.16} \\
\multicolumn{1}{c|}{} & \multicolumn{1}{l|}{\textcolor{black}{FFN-WavLM}} & \multicolumn{1}{l|}{\textcolor{black}{Raw Waveform}} & \textcolor{black}{38.99 $\pm$ 6.85} & \textcolor{black}{47.11 $\pm$ 5.67} & \textcolor{black}{34.96 $\pm$ 3.06} & \textcolor{black}{35.26 $\pm$ 3.67} & \textcolor{black}{30.99 $\pm$ 2.16} & \textcolor{black}{27.62 $\pm$ 2.19} \\ \hline
\multicolumn{1}{c|}{\multirow{6}{*}{\begin{tabular}[c]{@{}c@{}}Audio\\ Contrastive/\\ Non-\\ contrastive\end{tabular}}} & \multicolumn{1}{l|}{SSL-ACL} & \multicolumn{1}{l|}{Log-mel} & 71.97 $\pm$ 3.25 & 71.12 $\pm$ 0.19 & 72.31 $\pm$ 1.46 & 69.19 $\pm$ 0.33 & 69.71 $\pm$ 0.62 & 73.22 $\pm$ 3.89 \\
\multicolumn{1}{c|}{} & \multicolumn{1}{l|}{BYOL-A} & \multicolumn{1}{l|}{Log-mel} & 65.58 $\pm$ 6.50 & 68.18 $\pm$ 4.67 & 69.98 $\pm$ 3.27 & 70.89 $\pm$ 3.17 & 71.54 $\pm$ 5.39 & 71.42 $\pm$ 2.57 \\
\multicolumn{1}{c|}{} & \multicolumn{1}{l|}{COLA} & \multicolumn{1}{l|}{Log-mel} & 70.52 $\pm$ 0.43 & 71.77 $\pm$ 0.96 & 69.53 $\pm$ 0.73 & 70.17 $\pm$ 1.15 & 68.35 $\pm$ 2.00 & 61.79 $\pm$ 4.47 \\
\multicolumn{1}{c|}{} & \multicolumn{1}{l|}{MoCo} & \multicolumn{1}{l|}{MFCC} & 64.05 $\pm$ 6.27 & 64.70 $\pm$ 4.01 & 60.44 $\pm$ 3.76 & 58.80 $\pm$ 1.31 & 57.55 $\pm$ 2.64 & 56.40 $\pm$ 5.09 \\
\multicolumn{1}{c|}{} & \multicolumn{1}{l|}{SimCLR} & \multicolumn{1}{l|}{Log-mel} & 59.26 $\pm$ 4.26 & 67.58 $\pm$ 2.69 & 64.31 $\pm$ 4.00 & 60.55 $\pm$ 4.57 & 65.72 $\pm$ 0.53 & 63.75 $\pm$ 0.86 \\
\multicolumn{1}{c|}{} & \multicolumn{1}{l|}{VG-NCL-$\tau$} & \multicolumn{1}{l|}{Wav2Vec2} & 13.69 $\pm$ 1.98 & 12.79 $\pm$ 2.80 & 10.57 $\pm$ 2.26 & 14.62 $\pm$ 4.92 & 12.39 $\pm$ 4.28 & 18.26 $\pm$ 1.62 \\
\multicolumn{1}{c|}{} & \multicolumn{1}{l|}{VG-NCL-$\varsigma$} & \multicolumn{1}{l|}{Wav2Vec2} & 12.61 $\pm$ 1.50 & 22.94 $\pm$ 8.21 & 19.36 $\pm$ 6.91 & 13.17 $\pm$ 2.13 & 17.27 $\pm$ 2.81 & 16.97 $\pm$ 3.73 \\
\multicolumn{1}{c|}{} & \multicolumn{1}{l|}{\textbf{SIGNL}} & \multicolumn{1}{l|}{Wav2Vec2} & \textbf{10.16 $\pm$ 2.16} & \textbf{10.26 $\pm$ 0.97} & 9.87 $\pm$ 1.23 & 9.23 $\pm$ 1.73 & \textbf{8.37 $\pm$ 2.15} & \textbf{7.69 $\pm$ 1.04} \\ \hline
\end{tabular}
}
\end{table*}

\footnotetext[2]{\textcolor{black}{Compared methods include LCNN~\citep{wu2020light_LCNN}, AASIST~\citep{jung2022aasist_graph}, SENet~\citep{wang2024multi_senet}, FFN-WavLM~\citep{elkheir2025comprehensive_wavlm}, SSL-ACL~\citep{wang2023self_acl}, BYOL-A~\citep{niizumi2022byol_byola}, COLA~\citep{saeed2021contrastive_COLA}, MoCo~\citep{xia2021self_mocoEmbedding}, SimCLR~\citep{zhang2021contrastive_simclr}, VG-NCL derived from~\citep{guo2024architecture_graphclsurvey}.}}

\subsection{Baseline Models}
\label{sec:result}
We categorized the baseline models into (1) supervised models using both traditional and learnable audio features, and (2) self-supervised contrastive and non-contrastive models, including both graph and non-graph-based variants. This diverse selection of baselines enables a rigorous comparison of SIGNL’s contributions from multiple learning paradigms. 

The first group included supervised audio deepfake detection models, such as LCNN~\citep{wu2020light_LCNN}, which uses LFCC features with a CNN architecture, and AASIST~\citep{jung2022aasist_graph}, a graph-based model built on GAT-T~\citep{tak2021graph_GAT} and RawGAT-ST~\citep{tak2021end_rawGat} that processes raw audio directly. We also evaluated these supervised methods using learnable audio representations from SSL models, including Wav2Vec2~\citep{baevski2020wav2vec} and Whisper~\citep{radford2023robust_whisper}, which are fine-tuned for deepfake detection in recent state-of-the-art approaches~\citep{zhang2024audio_hybridlearnablevisual, kawa2023improved_whisper_AD}. Moreover, more recent methods, such as SENet~\citep{wang2024multi_senet} and FFN-WavLM~\citep{elkheir2025comprehensive_wavlm}, were also evaluated.

The second category consisted of recent contrastive and non-contrastive self-supervised learning methods for audio, such as COLA~\citep{saeed2021contrastive_COLA}, SimCLR~\citep{zhang2021contrastive_simclr}, BYOL-A~\citep{niizumi2022byol_byola}, MoCo~\citep{xia2021self_mocoEmbedding}, and SSL-ACL~\citep{wang2023self_acl}, which are not graph-based. Additionally, we evaluated VG-NCL, a vanilla graph-based non-contrastive learning method used as a benchmark for graph non-contrastive learning, following the design principles outlined by Guo et al.~\citep{guo2024architecture_graphclsurvey}, which was adapted to learn one graph per audio sample. For VG-NCL, we use either the temporal graph $G^{\tau}$ or the spectral graph $G^{\varsigma}$, both derived from learnable Wav2Vec2 representations to ensure a fair comparison with SIGNL. \textcolor{black}{All baselines were implemented using their official codebases with optimal hyperparameters and their preferred audio input types, including any augmentations specified in their original work. All baseline results reported in this paper were independently reproduced. No external results were cited directly from the literature, as all methods were evaluated under the limited-label data and cross-domain evaluation settings, which are part of our contribution.}

\subsection{(\textbf{RQ1}) Comparison with Baseline Methods}

\begin{table}[htbp]
\centering
\caption{Number of labels used in the training and development sets for ASVspoof 2021 DF, ASVspoof 5, and CFAD at different label percentages.}
\label{tb:labels}
\resizebox{\columnwidth}{!}{%
\begin{tabular}{r|cc|cc|cc}
\hline
\multicolumn{1}{c|}{\multirow{2}{*}{\textbf{\% Label}}} & \multicolumn{2}{c|}{\textbf{ASVspoof 2021 DF}} & \multicolumn{2}{c|}{\textbf{ASVspoof 5}} & \multicolumn{2}{c}{\textbf{CFAD}} \\ \cline{2-7} 
\multicolumn{1}{c|}{} & \textbf{Train} & \textbf{Dev} & \textbf{Train} & \textbf{Dev} & \textbf{Train} & \textbf{Dev} \\ \hline
5\% & 1,269 & 1,242 & 9,117 & 2,838 & 1,920 & 720 \\
10\% & 2,538 & 2,484 & 18,235 & 5,677 & 3,840 & 1,440 \\
30\% & 7,614 & 7,453 & 54,707 & 17,031 & 11,520 & 4,320 \\
50\% & 12,690 & 12,422 & 91,178 & 28,386 & 19,200 & 7,200 \\
80\% & 20,304 & 19,875 & 145,885 & 45,418 & 30,720 & 11,520 \\
Full & 25,380 & 24,844 & 182,357 & 56,773 & 38,400 & 14,400 \\ \hline
\end{tabular}
}
\end{table}

To address RQ1, each model was evaluated under varying levels of label availability: 5\%, 10\%, 30\%, 50\%, 80\%, and the full label set. The labeled subsets were applied exclusively to the training and development sets, while the test set always retained the complete dataset information. In the limited label scenarios, the original proportional distribution of bona fide and fake samples, as well as the diversity of attack types, was preserved to reflect the composition of the complete training set. This approach ensures consistency in sample selection, reduces potential biases, and facilitates fair comparisons across all baselines. The number of labels available for the train and dev sets is summarized in~\autoref{tb:labels}.

\noindent \textbf{In-domain Evaluation Results}. The results, summarized in~\autoref{tab:results}, show that SIGNL achieves the lowest EER across most label percentages and datasets: ASVspoof 2021 DF, ASVspoof 5, and CFAD. In the full-label scenario, SIGNL reaches mean EERs of 7.21\%, 2.33\%, and 8.44\% on ASVspoof 2021 DF, ASVspoof 5, and CFAD, respectively. While it achieves the best results on ASVspoof 2021 DF and ASVspoof 5, the best EER on CFAD (8.41\%) is achieved by FFN-WavLM.

Even with limited labeled data, SIGNL remains highly effective. With only 5\% of labeled data, it reaches mean EERs of 7.88\% on ASVspoof 2021 DF, 3.95\% on ASVspoof 5, and 9.90\% on CFAD, outperforming all baseline competitors on the ASVspoof datasets. On CFAD, however, VG-NCL-$\tau$ achieves a lower EER (9.48\%) under the same condition. Notably, SIGNL occasionally performs better with smaller labeled subsets, likely due to reduced labeled label noise and more consistent data quality, which can enhance representation learning.

SIGNL consistently outperforms supervised models like LCNN and AASIST, even when using similar front-end configurations, highlighting its ability to extract meaningful representations without heavy reliance on labeled data. By transforming audio into spectral-temporal graphs and applying diverse augmentations, SIGNL enhances input variability and effectively clusters positive samples, enabling robust feature learning during pre-training. This leads to stronger performance in low-label settings, where supervised models degrade significantly, for example, when only 5\% of labeled data is available.

SIGNL also surpasses contrastive and non-contrastive baselines such as SSL-ACL, BYOL-A, COLA, MoCo, and SimCLR, due to its graph-based modeling of complex spectral and temporal relationships. Unlike VG-NCL, which uses a single graph per sample, SIGNL captures interdependencies across both views, yielding a more comprehensive representation. This combination of spectral-temporal graph learning and non-contrastive pre-training enables strong detection performance, particularly under label-scarce conditions.

\noindent \textbf{Cross-domain Evaluation Results}. To evaluate generalization, we tested SIGNL and baselines on the In-The-Wild dataset, focusing on their ability to handle diverse audio deepfake attacks in out-of-domain scenarios. The results, shown in~\autoref{tab:contrast_itw}, reveal that SIGNL consistently demonstrates strong cross-domain generalization on the In-The-Wild dataset. For example, it achieved an EER of 10.89\% when trained on ASVspoof 2021 DF with full labels and 7.69\% when trained on CFAD, despite the different languages used for training and testing in this cross-domain setting. For this different language scenario, robust performance is also observed in other supervised methods using Wav2Vec2 as a front-end, likely due to Wav2Vec2's pre-training on multilingual data, which enhances its ability to produce learnable visual representations of audio.

In limited label scenarios, SIGNL performed well, achieving EERs of 14.85\% with 5\% labels on ASVspoof 2021 DF and 10.16\% with 5\% labels on CFAD. However, in some configurations, other methods performed competitively with or better than SIGNL. For instance, AASIST with Raw Waveform outperformed SIGNL when trained on ASVspoof 5 with 30\% and full label configurations. Similarly, AASIST with Wav2Vec2 as the front-end slightly outperformed SIGNL when trained on CFAD with 30\% label data.

The results indicate that \textit{cross-domain evaluation poses greater challenges than in-domain evaluation}, as evidenced by generally poorer performance on ASVspoof 5 across all baselines. This likely stems from the models not encountering the relevant characteristics of novel or unseen audio attack types during pre-training and downstream training, highlighting the difficulty of generalizing to \textit{out-of-distribution (OOD) attacks}. Further discussion of this issue is provided in~\autoref{sec:future}.

These findings underscore the importance of robust representation learning under low-resource settings. SIGNL’s performance advantage highlights the benefit of modeling both spectral and temporal dependencies explicitly. Its dual-view structure, when combined with graph augmentations and non-contrastive learning, offers significant gains in generalization, particularly when labeled data is scarce.

\subsection{(\textbf{RQ2}) Analysis on Graph Augmentations}
\label{sec:graph_aug}
To address RQ2, we evaluated all possible combinations of three graph augmentations: Edge Dropping (ED), Gaussian Noise (GN), and Feature Masking (FM) on the ASVspoof 2021 DF and ASVspoof 5 datasets. The parameters were configured as follows: an edge-dropping probability of 0.5, a Gaussian noise standard deviation of 0.1, and a feature masking probability of 0.5. We tested eight different experimental configurations, as outlined in~\autoref{tab:optimal_combinations_eval}. The results indicate that the SIGNL-8 variant, which combines all three augmentation techniques (ED, GN, and FM), achieved the best performance. Specifically, SIGNL-8 obtained the lowest EER scores, with 7.21\% on the ASVspoof 2021 DF, 2.33\% on the ASVspoof 5, and 8.44\% on the CFAD datasets. These findings underscore the importance of diverse graph augmentations in boosting representation learning quality during pre-training.

\begin{table}[!t]
\centering
\caption{EER (\%) comparison for SIGNL across different combinations of graph augmentations in the full-label scenario. All values represent mean $\pm$ standard deviation. Lower values: better performance ($\downarrow$). \textbf{Bold}: best result.}
\label{tab:optimal_combinations_eval}
\resizebox{\columnwidth}{!}{
\begin{tabular}{llllccl}
\hline
\textbf{Methods} & \textbf{ED} & \textbf{GN} & \textbf{FM} & \textbf{ASVspoof 2021 DF} & \textbf{ASVspoof 5} & \multicolumn{1}{c}{\textbf{CFAD}} \\ \hline
SIGNL-1 & -- & -- & -- & 8.57 $\pm$ 1.65 & 2.72 $\pm$ 0.94 & 8.63 $\pm$ 0.30 \\
SIGNL-2 & -- & -- & \checkmark &  9.19 $\pm$ 1.45 & 3.25 $\pm$ 1.46 &  8.80 $\pm$ 0.41\\
SIGNL-3 & -- & \checkmark & -- & 7.93 $\pm$ 0.54 & 2.59 $\pm$ 0.44 & 8.83 $\pm$ 0.39 \\
SIGNL-4 & -- & \checkmark & \checkmark & 8.80 $\pm$ 1.24 & 2.53 $\pm$ 1.07 & 8.64 $\pm$ 0.11 \\
SIGNL-5 & \checkmark & -- & -- & 8.02 $\pm$ 0.73 & 2.89 $\pm$ 1.13 & 8.58 $\pm$ 0.21 \\
SIGNL-6 & \checkmark & -- & \checkmark & 9.00 $\pm$ 2.52 & 2.42 $\pm$ 0.96 & 8.89 $\pm$ 0.47 \\
SIGNL-7 & \checkmark & \checkmark & -- & 9.40 $\pm$ 2.98 & 2.86 $\pm$ 0.75 & 8.57 $\pm$ 0.25 \\
SIGNL-8 & \checkmark & \checkmark & \checkmark & \textbf{7.21 $\pm$ 0.61} & \textbf{2.33 $\pm$ 0.66} & \textbf{8.44 $\pm$ 0.12} \\ \hline
\end{tabular}
}
\end{table}

\begin{figure*}[!ht]
  \centering
  \includegraphics[width=\textwidth]{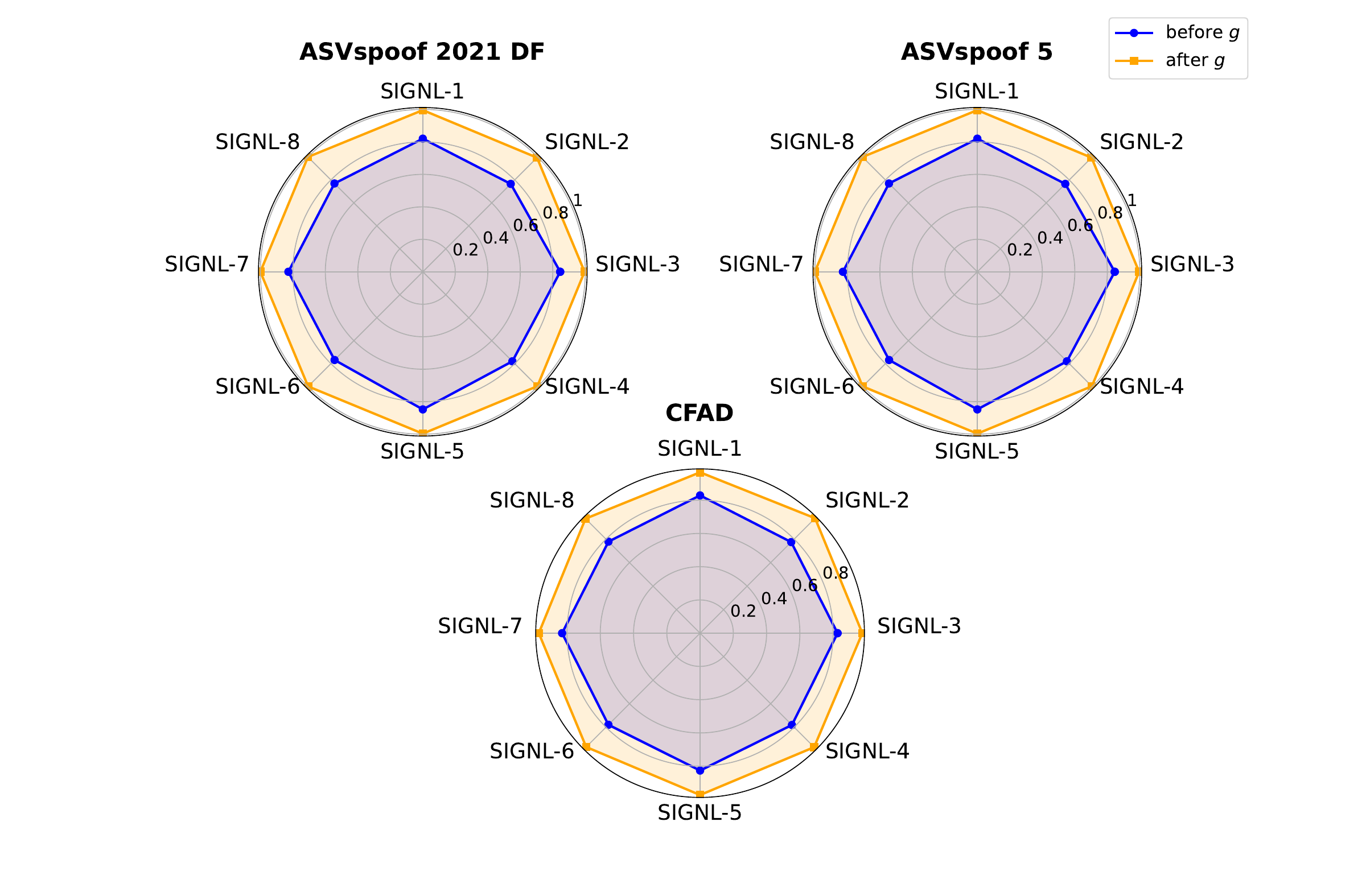}
  \caption{Similarity of the pair embeddings before and after the projection head $g$.}
  \label{img:collapse}
  \vspace{-2pt}
\end{figure*}

\noindent \textbf{Analysis on Feature Collapse}. A well-known challenge in contrastive learning is feature collapse, which occurs when the encoder generates identical features for all inputs, leading to a failure in learning meaningful representations~\citep{guo2024architecture_graphclsurvey}. This typically happens when the learning process relies solely on positive pairs without negative samples to introduce diversity. However, recent studies show that graph non-contrastive learning can avoid feature collapse in graph-level tasks, even without additional architectural modifications~\citep{guo2024architecture_graphclsurvey}. 

To validate this, we analyzed our model variants under different augmentation configurations by measuring the similarity of feature embeddings before and after passing through the projection head $g$. As shown in~\autoref{img:collapse}, SIGNL's various augmentation settings effectively prevent feature collapse. Although the projection head $g$ produces highly similar features (indicating a collapsed solution) with nearly perfect similarity scores, the similarity of the feature embeddings \textit{before projection} remains low, around 0.8 or less. Furthermore, the use of FM augmentation further reduces the similarity of feature embeddings, achieving scores closer to 0.7, as observed in variants SIGNL-2, SIGNL-4, SIGNL-6, and SIGNL-8. These overall pair similarity visualizations suggest that the encoders are effectively learning diverse and meaningful representations and are not suffering from collapse.

It is also important to note that the projection head is removed after pre-training, which is in line with common practices, ensuring that the learned representations are not affected by the collapsed outputs of the projection head during the downstream tasks.

\subsection{(\textbf{RQ3}) Ablation Studies}
\label{sec:ablation}
To address RQ3, we conducted ablation experiments by selectively removing specific components from our framework to evaluate their contributions. We explored three ablation settings: 
\begin{itemize}
    \item \textbf{Without Vision GC Encoders} (\textit{SIGNL w/o GNN}): Replacing vision GC encoders with CNN-based models, thus eliminating graph techniques.
    \item \textbf{Without Fine-tuning} (\textit{SIGNL w/o FT}): Using fixed encoder parameters without downstream fine-tuning.
    \item \textbf{Without Pre-training} (\textit{SIGNL w/o Pre}): Skipping the pre-training phase and directly performing supervised downstream training.
\end{itemize}

\begin{table}[!ht]
\centering
\caption{EER (\%) comparison for SIGNL under various ablation settings in the full-label scenario. All values represent mean $\pm$ standard deviation. Lower values: better performance ($\downarrow$). \textbf{Bold}: best result.}
\label{tab:ablation}
\resizebox{\columnwidth}{!}{%
\begin{tabular}{lccc}
\hline
\textbf{Methods} & \textbf{ASVspoof 2021 DF} & \textbf{ASVspoof 5} & \textbf{CFAD} \\ \hline
SIGNL w/o GNN & 13.28 $\pm$ 1.13 & 9.42 $\pm$ 2.93 & 10.59 $\pm$ 0.97 \\
SIGNL w/o FT & 9.23 $\pm$ 0.82 & 5.36 $\pm$ 1.67 & 15.18 $\pm$ 1.33 \\
SIGNL w/o Pre & 7.91 $\pm$ 0.73 & 2.61 $\pm$ 0.87 & 8.62 $\pm$ 1.14 \\ \hline
\textbf{SIGNL} & \textbf{7.21 $\pm$ 0.61} & \textbf{2.33 $\pm$ 0.66} & \textbf{8.44 $\pm$ 0.12} \\ \hline
\end{tabular}
}
\end{table}

\begin{table}[!ht]
\centering
\caption{EER (\%) comparison for SIGNL under various ablation settings in the limited-label scenario (5\% label). All values represent mean $\pm$ standard deviation. Lower values: better performance ($\downarrow$). \textbf{Bold}: best result.}
\label{tab:ablation_limited}
\resizebox{\columnwidth}{!}{%
\begin{tabular}{lccc}
\hline
\textbf{Methods} & \textbf{ASVspoof 2021 DF} & \textbf{ASVspoof 5} & \textbf{CFAD} \\ \hline
SIGNL w/o GNN & 12.05 $\pm$ 4.21 & 19.39 $\pm$ 4.73 & 15.18 $\pm$ 1.84 \\
SIGNL w/o FT & 38.00 $\pm$ 2.22 & 18.31 $\pm$ 1.00 & 31.51 $\pm$ 1.18 \\
SIGNL w/o Pre & 20.23 $\pm$ 5.91 & 6.71 $\pm$ 1.74 & 10.00 $\pm$ 1.33 \\ \hline
\textbf{SIGNL} & \textbf{7.88  $\pm$ 2.11} & \textbf{3.95 $\pm$ 1.34} & \textbf{9.90 $\pm$ 0.20} \\ \hline
\end{tabular}
}
\end{table}

\noindent \textcolor{black}{\textbf{Ablation Study Results}. The results are presented in~\autoref{tab:ablation} (full-label setting) and~\autoref{tab:ablation_limited} (limited-label setting). Across both settings, each component of SIGNL plays a critical role in performance:}
\begin{itemize}
    \item \textcolor{black}{\textbf{Impact of Removing GNNs.} Without the vision GC encoders (\textit{SIGNL w/o GNN}), EER increases significantly--for example, from 3.95\% to 19.39\% on ASVSpoof 5 in the limited-label setting.  This demonstrates the importance of graph modeling: while CNNs focus on local patterns, GNNs capture long-range dependencies and cross-patch relationships, which are essential for identifying subtle deepfake artifacts dispersed across the spectral-temporal structure.}

    \item \textcolor{black}{\textbf{Impact of Removing Fine-Tuning. } Without task-specific fine-tuning (\textit{SIGNL w/o FT}), the model's performance drops significantly, especially under label scarcity. For example, EER on CFAD increases from 9.90\% to 31.51\%. This highlights that even with a strong self-supervised backbone, downstream adaptation remains crucial to align learned features with task-specific decision boundaries.}

    \item \textcolor{black}{\textbf{Impact of Removing Pre-Training.} Removing the pre-training phase \textit{SIGNL w/o Pre} consistently degrades performance in low-lable scenarios--for example, increasing EER from 7.88\% to 20.23\% on ASVSpoof 2021 DF--demonstrating the importance of non-contrastive graph pre-training for extracting useful spectral-temporal features from unlabeled data. Interestingly, in the full-label setting, the absence of pre-training has only a marginal effect across all 3 datasets (e.g., EER increases from 7.21\% to 7.91\% on ASVSpoof 2021 DF), suggesting that pre-training is particularly valuable for low-resource conditions but less critical when abundant labeled data is available.}
\end{itemize}

\textcolor{black}{These results confirm that each component—vision GC encoders, fine-tuning, and non-contrastive pre-training—is crucial for achieving robust audio deepfake detection performance, particularly in limited-label scenarios.}

\begin{figure*}[!htb]
    \includegraphics[width=\textwidth]{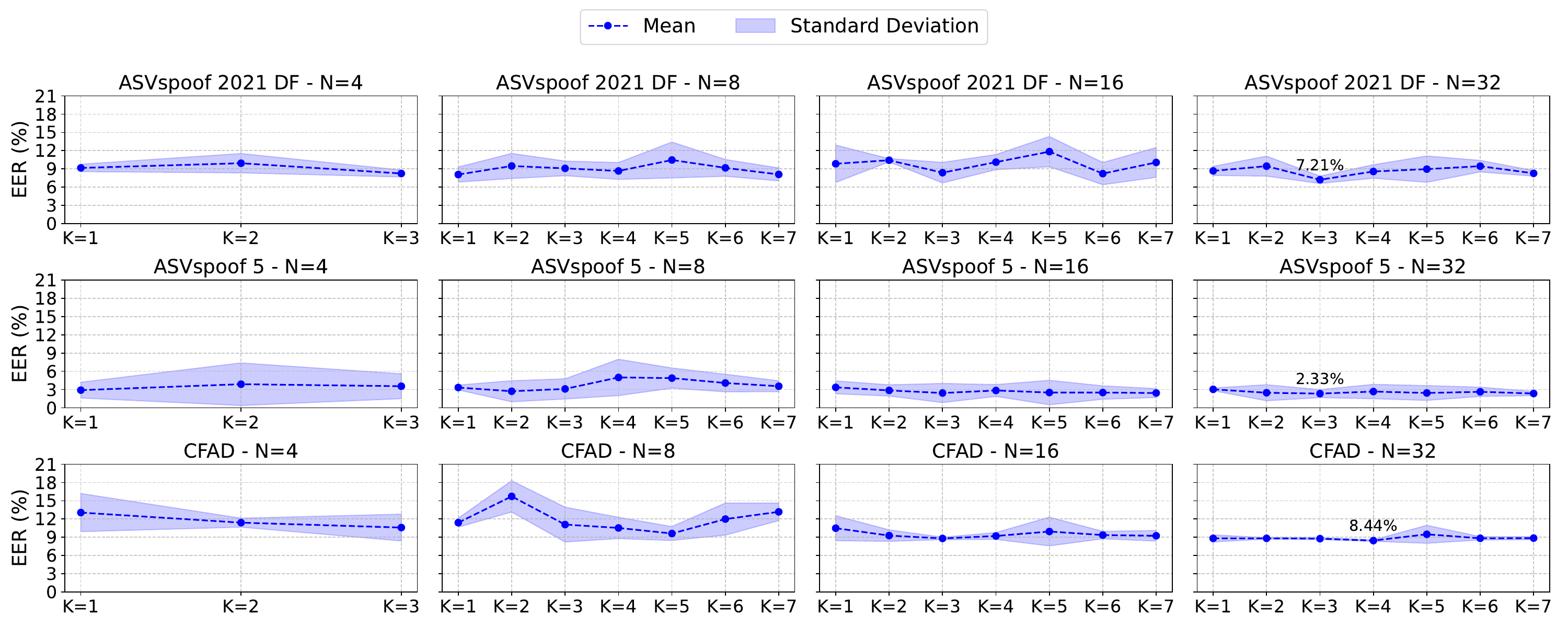}
    \caption{EER (\%) comparison for SIGNL across different combinations of the number of patches ($N$) and the number of nodes' neighbors ($K$) in the full-label scenario. The line plots represent the mean EER values, while the shaded areas indicate the standard deviation. Lower values: better performance ($\downarrow$).}
  \label{img:sens_analysis}
  \vspace{-2pt}
\end{figure*}

\subsection{(\textbf{RQ4}) Parameter Sensitivity}
\label{sec:sens}

\noindent\textbf{Sensitivity Analysis of Hyperparameters.} When converting the visual representation of audio into graph data, the number of patches, $N$, serves as a key hyperparameter that determines the number of nodes in both spectral and temporal graphs. To maintain consistency between spectral and temporal views, we use the same number of patches for both. We experimented with different patch sizes $N$ based on common divisors of Wav2Vec2's visual feature dimensions (1024, 224). The observed common divisors of 1024 and 224 are 4, 8, 16, and 32. In addition to patch size, we analyzed the impact of the number of neighbors $K$, another critical hyperparameter that controls the graph's connectivity by defining the number of relationships each node maintains. This parameter affects the graph's density, influencing its structure and characteristics. We tested $K$ values ranging from 1 to 7. As shown in~\autoref{img:sens_analysis}, the best results across all datasets were achieved with $N = 32$. Specifically, for the ASVspoof 2021 DF dataset, the lowest EER (7.21\%) was obtained with $K = 3$. Similarly, for ASVspoof 5, the best EER (2.33\%) was achieved with $K = 3$, and for CFAD, the lowest EER (8.44\%) occurred with $K = 4$. Larger $N$ values provide more detailed information that helps capture the local structure in the visual representation of audio. A value of $K = 3$ or $4$ may offer a balance by sharing enough information with neighbors without blurring distinct patterns.

\begin{figure}[!htb]
  \centering
  \includegraphics[width=\columnwidth]{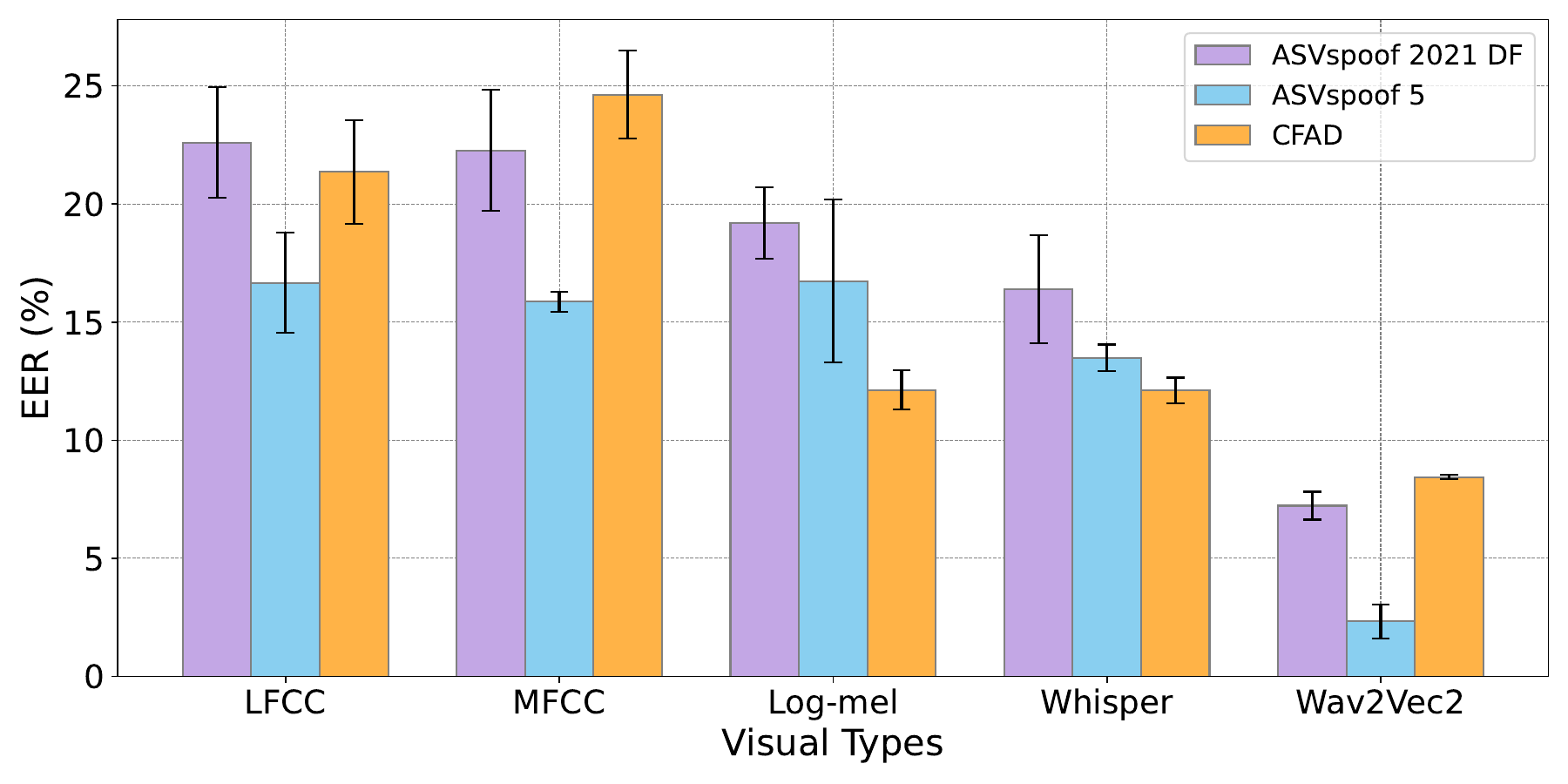}
  \caption{EER (\%) comparison for SIGNL across different visual representations of audio in the full-label scenario. The bar charts represent the mean, while the error bars indicate the standard deviation. Lower values: better performance ($\downarrow$). }
  \label{img:visrep}
  \vspace{-2pt}
\end{figure}

\noindent \textbf{Impact of Different Visual Represenations}. Converting raw audio into fixed-length representations, such as matrices, is crucial for information retrieval of audio data. This process captures the time-frequency characteristics of audio signals, enabling data mining techniques to analyze and extract meaningful patterns more effectively. These matrices can be visualized as heatmaps or other types of plots. We evaluated several visual representations, including Linear Frequency Cepstral Coefficients (LFCC), Mel-Frequency Cepstral Coefficients (MFCC), and Log-mel spectrogram. Additionally, we tested visual inputs derived from pre-trained SSL models, namely Whisper~\citep{radford2023robust_whisper} and Wav2Vec2~\citep{baevski2020wav2vec}. Whisper's 
"tiny" variant and Wav2Vec2's "wav2vec2-xls-r-300m" variant were used to generate learnable visual representations of audio data. As shown in~\autoref{img:visrep}, the Wav2Vec2 as visual representation delivered the best performance, yielding the lowest EER values across the datasets.

\subsection{Resilience to Input Perturbation}\label{adversarial}
In addition to evaluating performance across datasets and label regimes, we assess the system's resilience to input perturbations. Specifically, we apply the Fast Gradient Sign Method (FGSM)~\citep{goodfellow2014explaining_FGSM} with $\epsilon = \frac{8}{255}$ to simulate imperceptible changes in the input waveform. These perturbations affect the spectrogram and subsequently the constructed graph representations.

\begin{table}[!ht]
\centering
\caption{EER (\%) comparison for SIGNL under the FGSM adversarial attack in the limited-label scenario (5\% Label). All values represent mean $\pm$ standard deviation, with (+delta) indicating the increase in EER caused by the adversarial attack.}

\label{tab:adversarial}
\resizebox{\columnwidth}{!}{%
\begin{tabular}{l|cc}
\hline
\multicolumn{1}{c|}{\multirow{2}{*}{\textbf{Methods}}} & \multicolumn{2}{c}{\textbf{FGSM}, $\epsilon = \frac{8}{255}$} \\ \cline{2-3} 
\multicolumn{1}{c|}{} & \textbf{ASVspoof 2021 DF} & \textbf{CFAD} \\ \hline
\textcolor{black}{SSL-ACL~\citep{wang2023self_acl}} & 45.13 $\pm$ 1.69 (+9.4) & 46.66 $\pm$ 1.64 (+17.2) \\
\textcolor{black}{BYOL-A~\citep{niizumi2022byol_byola} } & 46.55 $\pm$ 4.91 (+23.0) & 48.70 $\pm$ 1.63 (+26.8) \\
\textcolor{black}{COLA~\citep{saeed2021contrastive_COLA}  }  & 37.40 $\pm$ 7.43 (+10.2) & 40.24 $\pm$ 3.29 (+19.1) \\
\textcolor{black}{MoCo~\citep{xia2021self_mocoEmbedding} }   & 29.27 $\pm$ 15.04 (+1.4) & 46.00 $\pm$ 8.51 (+17.4) \\
\textcolor{black}{SimCLR~\citep{zhang2021contrastive_simclr} } & 42.85 $\pm$ 5.57 (+17.9) & 53.06 $\pm$ 2.28 (+25.8) \\
\textcolor{black}{VG-NCL-$\tau$ derived from~\cite{guo2024architecture_graphclsurvey} }& 26.25 $\pm$ 3.38 (+12.0) & 13.85 $\pm$ 2.27 (+4.4) \\
\textcolor{black}{VG-NCL-$\varsigma$ derived from~\cite{guo2024architecture_graphclsurvey} }& 20.15 $\pm$ 4.98 (+10.8) & 14.99 $\pm$ 3.16 (+4.9) \\
\textbf{SIGNL} & \textbf{15.51 $\pm$ 0.84 (+7.6)} & \textbf{10.37 $\pm$ 1.56 (+0.5)} \\ \hline
\end{tabular}
}
\end{table}

As shown in~\autoref{tab:adversarial}, SIGNL maintains strong performance on CFAD with minimal degradation, and consistently outperforms all baselines under attack on both datasets. While performance declines more noticeably on ASVspoof 2021, SIGNL still remains the most robust among the evaluated models. These results indicate that SIGNL's graph-based structure contributes to stable representations, even in the presence of input noise or distortion.

\begin{table*}[!ht]
\centering
\caption{\textcolor{black}{EER (\%) comparison for zero-shot evaluation on CosyVoice 2.0 samples. The methods were trained under the limited-label scenario (5\% label). All values of the methods\protect\footnotemark[2] represent mean $\pm$ standard deviation. Lower values: better performance ($\downarrow$). \textbf{Bold}: best result.}}
\label{tab:cosyvoice_nofinetune}
\resizebox{\textwidth}{!}{%
\begin{tabular}{cllccc}
\hline
\multicolumn{1}{c|}{\multirow{2}{*}{\textcolor{black}{\textbf{Categories}}}} & 
\multicolumn{1}{c|}{\multirow{2}{*}{\textcolor{black}{\textbf{Methods}}}} & 
\multicolumn{1}{l|}{\multirow{2}{*}{\textcolor{black}{\textbf{Front-end}}}} & 
\multicolumn{3}{c}{\textcolor{black}{\textbf{Tested on CosyVoice 2.0 w/o FT}}} \\ \cline{4-6} 
\multicolumn{1}{c|}{} & 
\multicolumn{1}{c|}{} & 
\multicolumn{1}{l|}{} & 
\textcolor{black}{\textbf{Trained on ASVspoof2021 DF}} & 
\textcolor{black}{\textbf{Trained on ASVspoof 5}} & 
\textcolor{black}{\textbf{Trained on CFAD}} \\ \hline

\multicolumn{1}{c|}{\multirow{8}{*}{\textcolor{black}{Supervised}}} & \multicolumn{1}{l|}{\textcolor{black}{LCNN}} & \multicolumn{1}{l|}{\textcolor{black}{LFCC}} & \textcolor{black}{36.23 $\pm$ 9.05} & \textcolor{black}{27.54 $\pm$ 2.51} & \textcolor{black}{37.68 $\pm$ 2.51} \\
\multicolumn{1}{c|}{} & \multicolumn{1}{l|}{\textcolor{black}{LCNN}} & \multicolumn{1}{l|}{\textcolor{black}{Whisper}} & \textcolor{black}{52.17 $\pm$ 7.53} & \textcolor{black}{21.74 $\pm$ 0.00} & \textcolor{black}{44.93 $\pm$ 5.02} \\
\multicolumn{1}{c|}{} & \multicolumn{1}{l|}{\textcolor{black}{LCNN}} & \multicolumn{1}{l|}{\textcolor{black}{Wav2Vec2}} & \textcolor{black}{46.38 $\pm$ 6.64} & \textcolor{black}{49.28 $\pm$ 9.05} & \textcolor{black}{40.58 $\pm$ 2.51} \\
\multicolumn{1}{c|}{} & \multicolumn{1}{l|}{\textcolor{black}{AASIST}} & \multicolumn{1}{l|}{\textcolor{black}{Raw Waveform}} & \textcolor{black}{49.28 $\pm$ 5.02} & \textcolor{black}{23.19 $\pm$ 2.51} & \textcolor{black}{52.17 $\pm$ 13.04} \\
\multicolumn{1}{c|}{} & \multicolumn{1}{l|}{\textcolor{black}{AASIST}} & \multicolumn{1}{l|}{\textcolor{black}{Whisper}} & \textcolor{black}{52.17 $\pm$ 4.35} & \textcolor{black}{37.68 $\pm$ 15.27} & \textcolor{black}{66.67 $\pm$ 6.64} \\
\multicolumn{1}{c|}{} & \multicolumn{1}{l|}{\textcolor{black}{AASIST}} & \multicolumn{1}{l|}{\textcolor{black}{Wav2Vec2}} & \textcolor{black}{56.52 $\pm$ 8.70} & \textcolor{black}{43.48 $\pm$ 11.50} & \textcolor{black}{40.58 $\pm$ 9.05} \\
\multicolumn{1}{c|}{} & \multicolumn{1}{l|}{\textcolor{black}{SENet}} & \multicolumn{1}{l|}{\textcolor{black}{LFCC + MPE}} & \textcolor{black}{40.58 $\pm$ 5.02} & \textcolor{black}{\textbf{14.49 $\pm$ 2.51}} & \textcolor{black}{55.07 $\pm$ 2.51} \\
\multicolumn{1}{c|}{} & \multicolumn{1}{l|}{\textcolor{black}{FFN-WavLM}} & \multicolumn{1}{l|}{\textcolor{black}{Raw Waveform}} & \textcolor{black}{50.72 $\pm$ 6.64} & \textcolor{black}{24.64 $\pm$ 12.55} & \textcolor{black}{65.22 $\pm$ 0.00} \\ \hline

\multicolumn{1}{c|}{\multirow{8}{*}{\textcolor{black}{\begin{tabular}[c]{@{}c@{}}Audio\\ Contrastive/\\ Non-\\ contrastive\end{tabular}}}} & \multicolumn{1}{l|}{\textcolor{black}{SSL-ACL}} & \multicolumn{1}{l|}{\textcolor{black}{Log-mel}} & \textcolor{black}{49.28 $\pm$ 2.51} & \textcolor{black}{26.09 $\pm$ 4.35} & \textcolor{black}{39.13 $\pm$ 7.53} \\
\multicolumn{1}{c|}{} & \multicolumn{1}{l|}{\textcolor{black}{BYOL-A}} & \multicolumn{1}{l|}{\textcolor{black}{Log-mel}} & \textcolor{black}{37.68 $\pm$ 5.02} & \textcolor{black}{23.19 $\pm$ 5.02} & \textcolor{black}{39.13 $\pm$ 4.35} \\
\multicolumn{1}{c|}{} & \multicolumn{1}{l|}{\textcolor{black}{COLA}} & \multicolumn{1}{l|}{\textcolor{black}{Log-mel}} & \textcolor{black}{40.58 $\pm$ 2.51} & \textcolor{black}{21.74 $\pm$ 7.53} & \textcolor{black}{\textbf{34.78 $\pm$ 0.00}} \\
\multicolumn{1}{c|}{} & \multicolumn{1}{l|}{\textcolor{black}{MoCo}} & \multicolumn{1}{l|}{\textcolor{black}{MFCC}} & \textcolor{black}{\textbf{34.78 $\pm$ 4.35}} & \textcolor{black}{37.68 $\pm$ 6.64} & \textcolor{black}{40.58 $\pm$ 6.64} \\
\multicolumn{1}{c|}{} & \multicolumn{1}{l|}{\textcolor{black}{SimCLR}} & \multicolumn{1}{l|}{\textcolor{black}{Log-mel}} & \textcolor{black}{42.03 $\pm$ 2.51} & \textcolor{black}{21.74 $\pm$ 4.35} & \textcolor{black}{46.38 $\pm$ 5.02} \\
\multicolumn{1}{c|}{} & \multicolumn{1}{l|}{\textcolor{black}{VG-NCL-$\tau$}} & \multicolumn{1}{l|}{\textcolor{black}{Wav2Vec2}} & \textcolor{black}{55.07 $\pm$ 5.02} & \textcolor{black}{47.83 $\pm$ 4.35} & \textcolor{black}{46.38 $\pm$ 6.64} \\
\multicolumn{1}{c|}{} & \multicolumn{1}{l|}{\textcolor{black}{VG-NCL-$\varsigma$}} & \multicolumn{1}{l|}{\textcolor{black}{Wav2Vec2}} & \textcolor{black}{56.52 $\pm$ 4.35} & \textcolor{black}{44.93 $\pm$ 2.51} & \textcolor{black}{49.28 $\pm$ 6.64} \\
\multicolumn{1}{c|}{} & \multicolumn{1}{l|}{\textcolor{black}{\textbf{SIGNL}}} & \multicolumn{1}{l|}{\textcolor{black}{Wav2Vec2}} & \textcolor{black}{39.13 $\pm$ 4.35} & \textcolor{black}{37.68 $\pm$ 2.51} & \textcolor{black}{43.48 $\pm$ 4.35} \\ \hline
\end{tabular}%
}
\end{table*}

\setcounter{footnote}{2}
\subsection{\textcolor{black}{Ultra-realistic Fake Speech Evaluation}}
\label{sec:ultra}

\textcolor{black}{In addition to cross-domain evaluation on the In-The-Wild dataset, we assess model resilience against emerging ultra-realistic deepfakes using CosyVoice 2.0 samples~\citep{du2024cosyvoice}. We collected 187 official samples from the CosyVoice 2.0 website\footnote{\textcolor{black}{\url{https://funaudiollm.github.io/cosyvoice2/}.}}, including 23 real ("prompt") audios and 164 generated ("fake"). We evaluate two settings: \textbf{Zero-shot} and \textbf{Fine-tuning}.}

\noindent \textcolor{black}{\textbf{Zero-shot}. We directly evaluated pre-trained models--trained with 5\% labels on ASVspoof 2021 DF, ASVspoof 5, and CFAD)--on all 187 CosyVoice 2.0 samples. Results in Table~\ref{tab:cosyvoice_nofinetune} show a significant performance drop across all methods, highlighting the difficulty of generalizing to highly realistic, unseen attacks. CosyVoice 2.0 samples exhibit nuanced prosody and naturalistic cadence that differ substantially from synthetic artifacts seen during training, which likely contributes to this failure mode. These findings underscore the need for adaptation mechanisms (e.g., continual learning or fine-tuning) to maintain robustness as new audio deepfake methods emerge.}

\begin{table*}[!ht]
\centering
\caption{\textcolor{black}{EER (\%) comparison for fine-tuned models on CosyVoice 2.0 samples. The methods were initially trained under the limited-label scenario (5\% label) before finetuning with a combination of the previous dataset samples plus CosyVoice 2.0 samples. All values of the methods\protect\footnotemark[2] represent mean $\pm$ standard deviation. Lower values: better performance ($\downarrow$). \textbf{Bold}: best result.}}
\label{tab:cosyvoice_finetune}
\begin{tabular}{cllccc}
\hline
\multicolumn{1}{c|}{\multirow{2}{*}{\textcolor{black}{\textbf{Categories}}}} & 
\multicolumn{1}{c|}{\multirow{2}{*}{\textcolor{black}{\textbf{Methods}}}} & 
\multicolumn{1}{l|}{\multirow{2}{*}{\textcolor{black}{\textbf{Front-end}}}} & 
\multicolumn{3}{c}{\textcolor{black}{\textbf{Tested on CosyVoice 2.0}}} \\ \cline{4-6} 
\multicolumn{1}{c|}{} & 
\multicolumn{1}{c|}{} & 
\multicolumn{1}{l|}{} & 
\shortstack{\\[0em] \textcolor{black}{\textbf{Fine-tuned on}} \\ \textcolor{black}{\textbf{ASVspoof2021 DF's model}}} & 
\shortstack{\\[0em] \textcolor{black}{\textbf{Fine-tuned on}} \\ \textcolor{black}{\textbf{ASVspoof 5's Model}}} & 
\shortstack{\\[0em] \textcolor{black}{\textbf{Fine-tuned on}} \\ \textcolor{black}{\textbf{CFAD's model}}} \\ \hline

\multicolumn{1}{c|}{\multirow{8}{*}{\textcolor{black}{Supervised}}} & \multicolumn{1}{l|}{\textcolor{black}{LCNN}} & \multicolumn{1}{l|}{\textcolor{black}{LFCC}} & \textcolor{black}{33.33 $\pm$ 0.00} & \textcolor{black}{27.78 $\pm$ 4.81} & \textcolor{black}{33.33 $\pm$ 0.00} \\
\multicolumn{1}{c|}{} & \multicolumn{1}{l|}{\textcolor{black}{LCNN}} & \multicolumn{1}{l|}{\textcolor{black}{Whisper}} & \textcolor{black}{27.78 $\pm$ 4.81} & \textcolor{black}{33.33 $\pm$ 0.00} & \textcolor{black}{38.89 $\pm$ 4.81} \\
\multicolumn{1}{c|}{} & \multicolumn{1}{l|}{\textcolor{black}{LCNN}} & \multicolumn{1}{l|}{\textcolor{black}{Wav2Vec2}} & \textcolor{black}{30.56 $\pm$ 9.62} & \textcolor{black}{25.00 $\pm$ 8.33} & \textcolor{black}{27.78 $\pm$ 4.81} \\
\multicolumn{1}{c|}{} & \multicolumn{1}{l|}{\textcolor{black}{AASIST}} & \multicolumn{1}{l|}{\textcolor{black}{Raw Waveform}} & \textcolor{black}{30.56 $\pm$ 4.81} & \textcolor{black}{33.33 $\pm$ 8.33} & \textcolor{black}{36.11 $\pm$ 4.81} \\
\multicolumn{1}{c|}{} & \multicolumn{1}{l|}{\textcolor{black}{AASIST}} & \multicolumn{1}{l|}{\textcolor{black}{Whisper}} & \textcolor{black}{30.56 $\pm$ 4.81} & \textcolor{black}{25.00 $\pm$ 0.00} & \textcolor{black}{27.78 $\pm$ 4.81} \\
\multicolumn{1}{c|}{} & \multicolumn{1}{l|}{\textcolor{black}{AASIST}} & \multicolumn{1}{l|}{\textcolor{black}{Wav2Vec2}} & \textcolor{black}{22.22 $\pm$ 4.81} & \textcolor{black}{16.67 $\pm$ 0.00} & \textcolor{black}{16.67 $\pm$ 0.00} \\
\multicolumn{1}{c|}{} & \multicolumn{1}{l|}{\textcolor{black}{SENet}} & \multicolumn{1}{l|}{\textcolor{black}{LFCC + MPE}} & \textcolor{black}{25.00 $\pm$ 8.33} & \textcolor{black}{22.22 $\pm$ 4.81} & \textcolor{black}{30.56 $\pm$ 4.81} \\
\multicolumn{1}{c|}{} & \multicolumn{1}{l|}{\textcolor{black}{FFN-WavLM}} & \multicolumn{1}{l|}{\textcolor{black}{Raw Waveform}} & \textcolor{black}{33.33 $\pm$ 0.00} & \textcolor{black}{16.67 $\pm$ 0.00} & \textcolor{black}{22.22 $\pm$ 4.81} \\ \hline

\multicolumn{1}{c|}{\multirow{8}{*}{\textcolor{black}{\begin{tabular}[c]{@{}c@{}}Audio\\ Contrastive/\\ Non-\\ contrastive\end{tabular}}}} & \multicolumn{1}{l|}{\textcolor{black}{SSL-ACL}} & \multicolumn{1}{l|}{\textcolor{black}{Log-mel}} & \textcolor{black}{22.22 $\pm$ 4.81} & \textcolor{black}{13.89 $\pm$ 4.81} & \textcolor{black}{30.56 $\pm$ 4.81} \\
\multicolumn{1}{c|}{} & \multicolumn{1}{l|}{\textcolor{black}{BYOL-A}} & \multicolumn{1}{l|}{\textcolor{black}{Log-mel}} & \textcolor{black}{16.67 $\pm$ 0.00} & \textcolor{black}{13.89 $\pm$ 4.81} & \textcolor{black}{16.67 $\pm$ 0.00} \\
\multicolumn{1}{c|}{} & \multicolumn{1}{l|}{\textcolor{black}{COLA}} & \multicolumn{1}{l|}{\textcolor{black}{Log-mel}} & \textcolor{black}{22.22 $\pm$ 4.81} & \textcolor{black}{13.89 $\pm$ 4.81} & \textcolor{black}{25.00 $\pm$ 8.33} \\
\multicolumn{1}{c|}{} & \multicolumn{1}{l|}{\textcolor{black}{MoCo}} & \multicolumn{1}{l|}{\textcolor{black}{MFCC}} & \textcolor{black}{30.56 $\pm$ 4.81} & \textcolor{black}{\textbf{11.11 $\pm$ 4.81}} & \textcolor{black}{30.56 $\pm$ 9.62} \\
\multicolumn{1}{c|}{} & \multicolumn{1}{l|}{\textcolor{black}{SimCLR}} & \multicolumn{1}{l|}{\textcolor{black}{Log-mel}} & \textcolor{black}{25.00 $\pm$ 0.00} & \textcolor{black}{13.89 $\pm$ 4.81} & \textcolor{black}{19.44 $\pm$ 4.81} \\
\multicolumn{1}{c|}{} & \multicolumn{1}{l|}{\textcolor{black}{VG-NCL-$\tau$}} & \multicolumn{1}{l|}{\textcolor{black}{Wav2Vec2}} & \textcolor{black}{27.78 $\pm$ 4.81} & \textcolor{black}{19.44 $\pm$ 4.81} & \textcolor{black}{19.44 $\pm$ 4.81} \\
\multicolumn{1}{c|}{} & \multicolumn{1}{l|}{\textcolor{black}{VG-NCL-$\varsigma$}} & \multicolumn{1}{l|}{\textcolor{black}{Wav2Vec2}} & \textcolor{black}{16.67 $\pm$ 0.00} & \textcolor{black}{16.67 $\pm$ 8.33} & \textcolor{black}{\textbf{13.89 $\pm$ 4.81}} \\
\multicolumn{1}{c|}{} & \multicolumn{1}{l|}{\textcolor{black}{\textbf{SIGNL}}} & \multicolumn{1}{l|}{\textcolor{black}{Wav2Vec2}} & \textcolor{black}{\textbf{13.89 $\pm$ 4.81}} & \textcolor{black}{13.89 $\pm$ 4.81} & \textcolor{black}{\textbf{13.89 $\pm$ 4.81}} \\ \hline
\end{tabular}
\end{table*}

\noindent \textcolor{black}{\textbf{Fine-tuning}. Given the small size of the CosyVoice 2.0 dataset—even compared to the smallest 5\% label setting of the other datasets (i.e., 1,269 samples from ASVspoof 2021 DF)—we split the 187 samples evenly into 93 for fine-tuning and 94 for evaluation. To prevent \textit{catastrophic forgetting}, we supplemented the training set with 93 randomly selected samples from the original datasets, resulting in 186 training examples per model.
As a result, each model is fine-tuned on 186 audios and evaluated on 94 audios. Table~\ref{tab:cosyvoice_finetune} shows clear improvements over the zero-shot setting. This demonstrates that even limited fine-tuning can effectively adapt models to new attack distributions, provided that balancing strategies are used. SIGNL achieved the lowest average EER (13.89\%) on both the ASVspoof 2021 DF and CFAD pre-trained models, demonstrating strong adaptation to new attacks under limited supervision. While this consistency may partly reflect the small evaluation set and low number of real samples (12), it also indicates SIGNL's robustness to domain shifts when adaptation is possible. We expect performance could be further improved with a larger set of CosyVoice 2.0 samples. Nonetheless, SIGNL—like all models—struggled in the zero-shot setting, highlighting its vulnerability to OOD attacks. Future work could explore generating additional CosyVoice 2.0 samples using the official model. Even without labels, such data could be valuable for unsupervised pre-training using SIGNL's graph augmentation strategies. This may improve robustness to distributional shifts without requiring explicit fine-tuning.}

\subsection{\textcolor{black}{Limitations} and Future Directions}
\label{sec:future}
While SIGNL demonstrates strong performance under low-label and cross-domain settings, some limitations remain that merit further investigation. 

\textcolor{black}{\textbf{Generalization to out-of-distribution (OOD) attacks.} A key challenge in audio deepfake detection is maintaining performance on unseen or evolving attack types.  As shown in Table~\ref{tab:contrast_itw} (cross-domain generalization) and Table~\ref{tab:cosyvoice_nofinetune} (CosyVoice 2.0 zero-shot), SIGNL's performance degrades on OOD attacks that differ significantly from training data—particularly under zero-shot conditions. While SIGNL achieves top performance after few-shot fine-tuning, its drop in the zero-shot setting highlights a broader issue shared across detection methods: current models often overfit to known synthesis artifacts and struggle to generalize to novel manipulation techniques. The emergence of ultra-realistic methods (e.g., CosyVoice 2.0) exacerbates this, suggesting that even high-performing systems require continual adaptation to remain robust.}

\textcolor{black}{\textbf{Dependence on Unlabeled Data Quality and Diversity.} Although SIGNL is label-efficient, its success still hinges on access to sufficiently diverse and representative unlabeled data. In scenarios where both labeled and unlabeled data are scarce—such as with newly emerging languages, speakers, or attack models—pre-training may fail to produce robust encoders. This limitation is evident in our CosyVoice 2.0 experiment in Section~\ref{sec:ultra}, where limited sample availability constrained model performance.}

\textcolor{black}{\textbf{Sensitivity to Input Perturbations.}
SIGNL's performance relies on the integrity and alignment of spectral and temporal graph views derived from audio spectrograms. As such, it may be vulnerable to perturbations that subtly alter or misalign these graph structures. Although our adversarial robustness experiments (Section~\ref{adversarial}) show that SIGNL outperforms other models under FGSM perturbations, the observed degradation still highlights that it remains susceptible to input distortions. In particular, adversarial or semantically consistent perturbations that desynchronize the dual views may expose failure modes not yet captured by standard augmentations. Future work could explore view-consistency training, graph structure regularization, or perturbation-invariant encodings to strengthen SIGNL's robustness further.}

\textcolor{black}{\textbf{Future Directions.} To mitigate these limitations, we identify several promising research avenues:}
\begin{itemize}
    \item \textcolor{black}{\textbf{Domain adaptation and few-shot learning} could enable more effective generalization to unseen speakers, accents, or synthesis models~\citep{li2024graph_domainAdaptationFewshot, febrinanto2023graph_lifelong}.}
    \item \textcolor{black}{\textbf{Continual learning} approaches may allow SIGNL to incrementally adapt to new attack patterns without forgetting previously learned representations~\citep{febrinanto2025rehearsal, chen2025region_regularization}.}
    \item \textcolor{black}{\textbf{Graph-level regularization and consistency training} can enhance robustness to structured perturbations~\citep{sun2022adversarial_attack} and maintain alignment across spectral-temporal views.}
    \item \textcolor{black}{\textbf{Synthetic data augmentation and simulation} strategies tailored to diverse acoustic conditions and manipulations could enrich the training set and improve encoder generalization.}
\end{itemize}
\textcolor{black}{By addressing these challenges, future versions of SIGNL could offer more stable and adaptive performance in the face of increasingly sophisticated audio deepfakes.}

\section{Conclusion}
\label{sec:conclusion}
In this work, we presented SIGNL, a spectral-temporal vision graph non-contrastive learning framework for audio deepfake detection under limited-label conditions. SIGNL constructs paired spectral and temporal graphs from visual audio representations, capturing complementary frequency-time structures. Unlike prior single-view graph approaches, SIGNL jointly models both views and is pre-trained using label-free non-contrastive learning with diverse graph augmentations, followed by fine-tuning with minimal labeled data. \textcolor{black}{SIGNL is suitable for real-world applications as a countermeasure system, especially in situations where labeled data is limited but large amounts of unlabeled audio are available, such as call recordings, podcasts, and social media clips. Extensive experiments on four benchmarks (ASVspoof 2021 DF, ASVspoof 5, CFAD, and In-the-Wild) show that SIGNL consistently performs better than state-of-the-art graph and non-graph baselines in both in-domain and cross-domain settings. SIGNL achieves strong results using only 5\% labeled data, including 7.88\% EER on ASVspoof 2021 DF and 3.95\% EER on ASVspoof 5, surpassing all baselines and showing its strength in low-label settings. It also generalizes well to unseen conditions, reaching 10.16\% EER on the In-the-Wild dataset when trained on CFAD.}

\section*{Acknowledgements}
The authors would like to thank Professor Feng Xia (RMIT University) for his supervision and valuable advice during the early stages of this work.

\bibliographystyle{apalike}
\bibliography{bib}

@String{Computing = "Computing" }

@article{yi2023audio_audioDeepfakeSurvey,
  title={Audio Deepfake Detection: A Survey},
  author={Yi, Jiangyan and Wang, Chenglong and Tao, Jianhua and Zhang, Xiaohui and Zhang, Chu Yuan and Zhao, Yan},
  journal={arXiv preprint arXiv:2308.14970},
  year={2023}
}

@inproceedings{choi2024dddm_vc,
  title={Dddm-vc: Decoupled denoising diffusion models with disentangled representation and prior mixup for verified robust voice conversion},
  author={Choi, Ha-Yeong and Lee, Sang-Hoon and Lee, Seong-Whan},
  booktitle={AAAI},
  year={2024}
}

@inproceedings{popov2021diffusion_vc,
  title={Diffusion-Based Voice Conversion with Fast Maximum Likelihood Sampling Scheme},
  author={Popov, Vadim and Vovk, Ivan and Gogoryan, Vladimir and Sadekova, Tasnima and Kudinov, Mikhail Sergeevich and Wei, Jiansheng},
  booktitle={ICLR},
  year={2021}
}

@inproceedings{tak2021end_cnnbased,
  title={End-to-end anti-spoofing with rawnet2},
  author={Tak, Hemlata and Patino, Jose and Todisco, Massimiliano and Nautsch, Andreas and Evans, Nicholas and Larcher, Anthony},
  booktitle={ICASSP},
  year={2021},
}

@inproceedings{wu2020light_LCNN,
  title={Light convolutional neural network with feature genuinization for detection of synthetic speech attacks},
  author={Wu, Zhenzong and Das, Rohan Kumar and Yang, Jichen and Li, Haizhou},
  booktitle={INTERSPEECH},
  year={2020}
}

@article{chen2021pindrop_resnet,
  title={Pindrop labs’ submission to the ASVspoof 2021 challenge},
  author={Chen, Tianxiang and Khoury, Elie and Phatak, Kedar and Sivaraman, Ganesh},
  journal={Proc. 2021 Edition of the Automatic Speaker Verification and Spoofing Countermeasures Challenge},
  pages={89--93},
  year={2021}
}

@inproceedings{liu2023leveraging_transformer,
  title={Leveraging positional-related local-global dependency for synthetic speech detection},
  author={Liu, Xiaohui and Liu, Meng and Wang, Longbiao and Lee, Kong Aik and Zhang, Hanyi and Dang, Jianwu},
  booktitle={ICASSP},
  year={2023},
}

@inproceedings{chen2023graph_graph,
  title={Graph-based spectro-temporal dependency modeling for anti-spoofing},
  author={Chen, Feng and Deng, Shiwen and Zheng, Tieran and He, Yongjun and Han, Jiqing},
  booktitle={ICASSP},
  year={2023},
}

@inproceedings{jung2022aasist_graph,
  title={Aasist: Audio anti-spoofing using integrated spectro-temporal graph attention networks},
  author={Jung, Jee-weon and Heo, Hee-Soo and Tak, Hemlata and Shim, Hye-jin and Chung, Joon Son and Lee, Bong-Jin and Yu, Ha-Jin and Evans, Nicholas},
  booktitle={ICASSP},
  year={2022},
}

@article{han2022vision,
  title={Vision gnn: An image is worth graph of nodes},
  author={Han, Kai and Wang, Yunhe and Guo, Jianyuan and Tang, Yehui and Wu, Enhua},
  journal={NeurIPS},
  year={2022}
}

@article{ericsson2022self_SSLsurvey,
  title={Self-supervised representation learning: Introduction, advances, and challenges},
  author={Ericsson, Linus and Gouk, Henry and Loy, Chen Change and Hospedales, Timothy M},
  journal={IEEE Signal Processing Magazine},
  volume={39},
  number={3},
  pages={42--62},
  year={2022},
  publisher={IEEE}
}

@inproceedings{he2020momentum_MoCo,
  title={Momentum contrast for unsupervised visual representation learning},
  author={He, Kaiming and Fan, Haoqi and Wu, Yuxin and Xie, Saining and Girshick, Ross},
  booktitle={CVPR},
  year={2020}
}

@inproceedings{chen2020simple_simclr,
  title={A simple framework for contrastive learning of visual representations},
  author={Chen, Ting and Kornblith, Simon and Norouzi, Mohammad and Hinton, Geoffrey},
  booktitle={ICML},
  year={2020}
}

@inproceedings{chen2021exploring_siamese,
  title={Exploring simple siamese representation learning},
  author={Chen, Xinlei and He, Kaiming},
  booktitle={CVPR},
  year={2021}
}

@article{grill2020bootstrap_byol,
  title={Bootstrap your own latent-a new approach to self-supervised learning},
  author={Grill, Jean-Bastien and Strub, Florian and Altch{\'e}, Florent and Tallec, Corentin and Richemond, Pierre and Buchatskaya, Elena and Doersch, Carl and Avila Pires, Bernardo and Guo, Zhaohan and Gheshlaghi Azar, Mohammad and others},
  journal={NeurIPS},
  year={2020}
}

@article{guo2024architecture_graphclsurvey,
  title={Architecture matters: Uncovering implicit mechanisms in graph contrastive learning},
  author={Guo, Xiaojun and Wang, Yifei and Wei, Zeming and Wang, Yisen},
  journal={NeurIPS},
  year={2024}
}

@inproceedings{tak2021end_rawGat,
  title={End-to-End Spectro-Temporal Graph Attention Networks for Speaker Verification Anti-Spoofing and Speech Deepfake Detection},
  author={Tak, Hemlata and Jung, Jee-Weon and Patino, Jose and Kamble, Madhu and Todisco, Massimiliano and Evans, Nicholas},
  booktitle={ASVspoof Workshop},
  year={2021}
}

@inproceedings{tak2021graph_GAT,
  title={Graph attention networks for anti-spoofing},
  author={Tak, Hemlata and Jung, Jee-weon and Patino, Jose and Todisco, Massimiliano and Evans, Nicholas},
  booktitle={INTERSPEECH},
  year={2021}
}

@inproceedings{velivckovic2018graph_GAT,
  title={Graph Attention Networks},
  author={Veli{\v{c}}kovi{\'c}, Petar and Cucurull, Guillem and Casanova, Arantxa and Romero, Adriana and Li{\`o}, Pietro and Bengio, Yoshua},
  booktitle={ICLR},
  year={2018}
}

@inproceedings{xia2021self_mocoEmbedding,
  title={Self-supervised text-independent speaker verification using prototypical momentum contrastive learning},
  author={Xia, Wei and Zhang, Chunlei and Weng, Chao and Yu, Meng and Yu, Dong},
  booktitle={ICASSP},
  year={2021}
}

@inproceedings{saeed2021contrastive_COLA,
  title={Contrastive learning of general-purpose audio representations},
  author={Saeed, Aaqib and Grangier, David and Zeghidour, Neil},
  booktitle={ICASSP},
  year={2021},
}

@inproceedings{wang2023self_acl,
  title={Self-supervised learning of audio representations using angular contrastive loss},
  author={Wang, Shanshan and Tripathy, Soumya and Mesaros, Annamaria},
  booktitle={ICASSP},
  year={2023},
}

@inproceedings{welling2016semi_gcn,
  title={Semi-supervised classification with graph convolutional networks},
  author={Welling, Max and Kipf, Thomas N},
  booktitle={ICLR},
  year={2017}
}

@inproceedings{he2016deep_restnetori,
  title={Deep residual learning for image recognition},
  author={He, Kaiming and Zhang, Xiangyu and Ren, Shaoqing and Sun, Jian},
  booktitle={CVPR},
  year={2016}
}

@article{oord2018representation_infoNCE,
  title={Representation learning with contrastive predictive coding},
  author={Oord, Aaron van den and Li, Yazhe and Vinyals, Oriol},
  journal={arXiv preprint arXiv:1807.03748},
  year={2018}
}

@inproceedings{todisco2019asvspoof_asvspoof2019,
  title={ASVspoof 2019: Future horizons in spoofed and fake audio detection},
  author={Todisco, Massimiliano and Wang, Xin and Vestman, Ville and Sahidullah, Md and Delgado, H{\'e}ctor and Nautsch, Andreas and Yamagishi, Junichi and Evans, Nicholas and Kinnunen, Tomi and Lee, Kong Aik},
  booktitle={INTERSPEECH},
  year={2019}
}

@inproceedings{yamagishi2021asvspoof_21,
  title={ASVspoof 2021: accelerating progress in spoofed and deepfake speech detection},
  author={Yamagishi, Junichi and Wang, Xin and Todisco, Massimiliano and Sahidullah, Md and Patino, Jose and Nautsch, Andreas and Liu, Xuechen and Lee, Kong Aik and Kinnunen, Tomi and Evans, Nicholas and others},
  booktitle={ASVspoof Workshop},
  year={2021}
}

@inproceedings{muller2022does_itw,
  title={Does audio deepfake detection generalize?},
  author={M{\"u}ller, Nicolas M and Czempin, Pavel and Dieckmann, Franziska and Froghyar, Adam and B{\"o}ttinger, Konstantin},
  booktitle={INTERSPEECH},
  year={2022}
}

@article{niizumi2022byol_byola,
  title={BYOL for audio: Exploring pre-trained general-purpose audio representations},
  author={Niizumi, Daisuke and Takeuchi, Daiki and Ohishi, Yasunori and Harada, Noboru and Kashino, Kunio},
  journal={IEEE/ACM Transactions on Audio, Speech, and Language Processing},
  volume={31},
  pages={137--151},
  year={2022},
  publisher={IEEE}
}

@inproceedings{zhang2021contrastive_simclr,
  title={Contrastive self-supervised learning for text-independent speaker verification},
  author={Zhang, Haoran and Zou, Yuexian and Wang, Helin},
  booktitle={ICASSP},
  year={2021}
}

@inproceedings{radford2023robust_whisper,
  title={Robust speech recognition via large-scale weak supervision},
  author={Radford, Alec and Kim, Jong Wook and Xu, Tao and Brockman, Greg and McLeavey, Christine and Sutskever, Ilya},
  booktitle={ICLR},
  year={2023},
}

@article{febrinanto2023graph_lifelong,
  title={Graph lifelong learning: A survey},
  author={Febrinanto, Falih Gozi and Xia, Feng and Moore, Kristen and Thapa, Chandra and Aggarwal, Charu},
  journal={IEEE Computational Intelligence Magazine},
  volume={18},
  number={1},
  pages={32--51},
  year={2023},
  publisher={IEEE}
}

@article{liu2022graph_graphssl,
  title={Graph self-supervised learning: A survey},
  author={Liu, Yixin and Jin, Ming and Pan, Shirui and Zhou, Chuan and Zheng, Yu and Xia, Feng and Philip, S Yu},
  journal={IEEE transactions on knowledge and data engineering},
  volume={35},
  number={6},
  pages={5879--5900},
  year={2022},
  publisher={IEEE}
}

@article{ding2022data_graphaug,
  title={Data augmentation for deep graph learning: A survey},
  author={Ding, Kaize and Xu, Zhe and Tong, Hanghang and Liu, Huan},
  journal={ACM SIGKDD Explorations Newsletter},
  volume={24},
  number={2},
  pages={61--77},
  year={2022},
  publisher={ACM New York, NY, USA}
}

@article{ju2024towards_gcl,
  title={Towards Graph Contrastive Learning: A Survey and Beyond},
  author={Ju, Wei and Wang, Yifan and Qin, Yifang and Mao, Zhengyang and Xiao, Zhiping and Luo, Junyu and Yang, Junwei and Gu, Yiyang and Wang, Dongjie and Long, Qingqing and others},
  journal={arXiv preprint arXiv:2405.11868},
  year={2024}
}

@inproceedings{liu2022discovering_falsenegativepushapart,
  title={Discovering informative and robust positives for video domain adaptation},
  author={Liu, Chang and Li, Kunpeng and Stopa, Michael and Amano, Jun and Fu, Yun},
  booktitle={ICLR},
  year={2022}
}

@article{baevski2020wav2vec,
  title={wav2vec 2.0: A framework for self-supervised learning of speech representations},
  author={Baevski, Alexei and Zhou, Yuhao and Mohamed, Abdelrahman and Auli, Michael},
  journal={NeurIPS},
  year={2020}
}

@article{wang2024asvspoof_asvspoof5,
  title={ASVspoof 5: Crowdsourced speech data, deepfakes, and adversarial attacks at scale},
  author={Wang, Xin and Delgado, Hector and Tak, Hemlata and Jung, Jee-weon and Shim, Hye-jin and Todisco, Massimiliano and Kukanov, Ivan and Liu, Xuechen and Sahidullah, Md and Kinnunen, Tomi and others},
  journal={arXiv preprint arXiv:2408.08739},
  year={2024}
}

@article{ma2024cfad,
  title={CFAD: A Chinese dataset for fake audio detection},
  author={Ma, Haoxin and Yi, Jiangyan and Wang, Chenglong and Yan, Xinrui and Tao, Jianhua and Wang, Tao and Wang, Shiming and Fu, Ruibo},
  journal={Speech Communication},
  volume={164},
  pages={103122},
  year={2024},
  publisher={Elsevier}
}

@article{tan2024naturalspeech_TTS,
  title={Naturalspeech: End-to-end text-to-speech synthesis with human-level quality},
  author={Tan, Xu and Chen, Jiawei and Liu, Haohe and Cong, Jian and Zhang, Chen and Liu, Yanqing and Wang, Xi and Leng, Yichong and Yi, Yuanhao and He, Lei and others},
  journal={IEEE Transactions on Pattern Analysis and Machine Intelligence},
  year={2024},
  publisher={IEEE}
}

@inproceedings{wang2024spikevoice_TTS,
  title={SpikeVoice: High-Quality Text-to-Speech Via Efficient Spiking Neural Network},
  author={Wang, Kexin and Zhang, Jiahong and Ren, Yong and Yao, Man and Shang, Di and Xu, Bo and Li, Guoqi},
  booktitle={Proceedings of the 62nd Annual Meeting of the Association for Computational Linguistics (Volume 1: Long Papers)},
  pages={7927--7940},
  year={2024}
}

@inproceedings{tak2022wav2vec_aasist,
  title={Automatic speaker verification spoofing and deepfake detection using wav2vec 2.0 and data augmentation},
  author={Tak, Hemlata and Todisco, Massimiliano and Wang, Xin and Jung, Jee-weon and Yamagishi, Junichi and Evans, Nicholas},
  booktitle    = {Proc. The Speaker and Language Recognition Workshop (Odyssey)},
  year         = {2022},
}

@article{kawa2023improved_whisper_AD,
  title={Improved deepfake detection using whisper features},
  author={Kawa, Piotr and Plata, Marcin and Czuba, Micha{\l} and Szyma{\'n}ski, Piotr and Syga, Piotr},
  booktitle={INTERSPEECH},
  year={2023}
}

@inproceedings{zhang2024audio_hybridlearnablevisual,
  title={Audio deepfake detection with self-supervised XLS-R and SLS classifier},
  author={Zhang, Qishan and Wen, Shuangbing and Hu, Tao},
  booktitle={Proceedings of the 32nd ACM International Conference on Multimedia},
  pages={6765--6773},
  year={2024}
}

@inproceedings{li2024graph_domainAdaptationFewshot,
  title={Graph Anomaly Detection with Domain-Agnostic Pre-Training and Few-Shot Adaptation},
  author={Li, Xujia and Chen, Lei},
  booktitle={ICDE},
  year={2024},
  organization={IEEE}
}

@inproceedings{thakoorlarge_bgrl,
	title        = {Large-Scale Representation Learning on Graphs via Bootstrapping},
	author       = {Thakoor, Shantanu and Tallec, Corentin and Azar, Mohammad Gheshlaghi and Azabou, Mehdi and Dyer, Eva L and Munos, Remi and Veli{\v{c}}kovi{\'c}, Petar and Valko, Michal},
	year         = 2022,
	booktitle    = {International Conference on Learning Representations}
}

@article{zhang2021canonical_CCASSG,
	title        = {From canonical correlation analysis to self-supervised graph neural networks},
	author       = {Zhang, Hengrui and Wu, Qitian and Yan, Junchi and Wipf, David and Yu, Philip S},
	year         = 2021,
	journal      = {Advances in Neural Information Processing Systems},
	volume       = 34,
	pages        = {76--89}
}

@article{bielak2022graph_GBT,
	title        = {Graph barlow twins: A self-supervised representation learning framework for graphs},
	author       = {Bielak, Piotr and Kajdanowicz, Tomasz and Chawla, Nitesh V},
	year         = 2022,
	journal      = {Knowledge-Based Systems},
	publisher    = {Elsevier},
	volume       = 256,
	pages        = 109631
}

@inproceedings{goodfellow2014explaining_FGSM,
  title={Explaining and harnessing adversarial examples},
  author={Goodfellow, Ian J and Shlens, Jonathon and Szegedy, Christian},
  booktitle={ICLR},
  year={2015}
}

@article{xia2025graph,
  title={Graph Learning},
  author={Xia, Feng and Peng, Ciyuan and Ren, Jing and Febrinanto, Falih Gozi and Luo, Renqiang and Saikrishna, Vidya and Yuo, Shuo and Kong, Xiangjie},
  journal={Foundations and Trends{\textregistered} in Signal Processing},
  volume={19},
  number={4},
  pages={371--551},
  year={2025},
  publisher={Now Publishers Boston—Delft}
}

@inproceedings{elkheir2025comprehensive_wavlm,
  title={Comprehensive Layer-Wise Analysis of SSL Models for Audio Deepfake Detection},
  author={Elkheir, Yassine and Samih, Younes and Maharjan, Suraj and Polzehl, Tim and Moeller, Sebastian},
  booktitle={Findings of the Association for Computational Linguistics: NAACL 2025},
  year={2025}
}

@inproceedings{wang2024multi_senet,
  title={Multi-scale permutation entropy for audio deepfake detection},
  author={Wang, Chenglong and He, Jiayi and Yi, Jiangyan and Tao, Jianhua and Zhang, Chu Yuan and Zhang, Xiaohui},
  booktitle={ICASSP},
  year={2024},
}

@inproceedings{febrinanto2025rehearsal,
  title={Rehearsal with Auxiliary-Informed Sampling for Audio Deepfake Detection},
  author={Febrinanto, Falih Gozi and Moore, Kristen and Thapa, Chandra and Ma, Jiangang and Saikrishna, Vidya and Xia, Feng},
  booktitle={INTERSPEECH},
  year={2025}
}

@article{du2024cosyvoice,
  title={Cosyvoice 2: Scalable streaming speech synthesis with large language models},
  author={Du, Zhihao and Wang, Yuxuan and Chen, Qian and Shi, Xian and Lv, Xiang and Zhao, Tianyu and Gao, Zhifu and Yang, Yexin and Gao, Changfeng and Wang, Hui and others},
  journal={arXiv preprint arXiv:2412.10117},
  year={2024}
}

@inproceedings{chen2025region_regularization,
  title={Region-based optimization in continual learning for audio deepfake detection},
  author={Chen, Yujie and Yi, Jiangyan and Fan, Cunhang and Tao, Jianhua and Ren, Yong and Zeng, Siding and Zhang, Chu Yuan and Yan, Xinrui and Gu, Hao and Xue, Jun and others},
  booktitle={AAAI},
  year={2025}
}

@article{sun2022adversarial_attack,
  title={Adversarial attack and defense on graph data: A survey},
  author={Sun, Lichao and Dou, Yingtong and Yang, Carl and Zhang, Kai and Wang, Ji and Yu, Philip S and He, Lifang and Li, Bo},
  journal={IEEE Transactions on Knowledge and Data Engineering},
  volume={35},
  number={8},
  pages={7693--7711},
  year={2022},
  publisher={IEEE}
}

@article{li2024guest,
  title={Guest editorial: Deep neural networks for graphs: Theory, models, algorithms, and applications},
  author={Li, Ming and Micheli, Alessio and Wang, Yu Guang and Pan, Shirui and Li{\'o}, Pietro and Gnecco, Giorgio Stefano and Sanguineti, Marcello},
  journal={IEEE Transactions on Neural Networks and Learning Systems},
  volume={35},
  number={4},
  pages={4367--4372},
  year={2024},
  publisher={IEEE}
}

@article{zheng2022graph,
  title={Graph neural networks for graphs with heterophily: A survey},
  author={Zheng, Xin and Wang, Yi and Liu, Yixin and Li, Ming and Zhang, Miao and Jin, Di and Yu, Philip S and Pan, Shirui},
  journal={arXiv preprint arXiv:2202.07082},
  year={2024}
}

@article{li2024permutation,
  title={Permutation equivariant graph framelets for heterophilous graph learning},
  author={Li, Jianfei and Zheng, Ruigang and Feng, Han and Li, Ming and Zhuang, Xiaosheng},
  journal={IEEE Transactions on neural networks and learning systems},
  volume={35},
  number={9},
  pages={11634--11648},
  year={2024},
  publisher={IEEE}
}

@article{zhou2025data,
  title={Data augmentation on graphs: A technical survey},
  author={Zhou, Jiajun and Xie, Chenxuan and Gong, Shengbo and Wen, Zhenyu and Zhao, Xiangyu and Xuan, Qi and Yang, Xiaoniu},
  journal={ACM Computing Surveys},
  volume={57},
  number={11},
  pages={1--34},
  year={2025},
  publisher={ACM New York, NY}
}

\end{document}